\documentclass[openacc]{rsproca_new}

\usepackage{soul,makecell,float}
\begin{document}

\title{Economists' erroneous estimates of damages from climate change}


\author{
Steve Keen$^{1}$, Timothy M. Lenton$^{2}$, Antoine Godin$^{3}$, Devrim Yilmaz$^{4}$, Matheus Grasselli$^{5}$, and Timothy J. Garrett$^{6}$}

\address{$^{1}$University College London, \href{url}{s.keen@isrs.org.uk}\\
$^{2}$University of Exeter, \href{url}{t.m.lenton@exeter.ac.uk}\\
$^{3}$Agence Francaise de Developpement, \href{url}{godina@afd.fr}\\
$^{4}$Agence Francaise de Developpement, \href{url}{yilmazsd@afd.fr}\\
$^{5}$McMaster University, \href{url}{grasselli@math.mcmaster.ca}\\
$^{6}$University of Utah, \href{url}{tim.garrett@utah.edu}}

\begin{abstract}
Economists have predicted that damages from global warming will be as low as 2.1\% of global economic production for a 3$^\circ$C rise in global average surface temperature, and 7.9\% for a 6$^\circ$C rise. Such relatively trivial estimates of economic damages---when these economists otherwise assume that human economic productivity will be an order of magnitude higher than today---contrast strongly with predictions made by scientists of significantly reduced human habitability from climate change. Nonetheless, the coupled economic and climate models used to make such predictions have been influential in the international climate change debate and policy prescriptions. Here we review the empirical work done by economists and show that it severely underestimates damages from climate change by committing several methodological errors, including neglecting tipping points, and assuming that economic sectors not exposed to the weather are insulated from climate change. Most fundamentally, the influential Integrated Assessment Model DICE is shown to be incapable of generating an economic collapse, regardless of the level of damages. Given these flaws, economists' empirical estimates of economic damages from global warming should be rejected as unscientific, and models that have been calibrated to them, such as DICE, should not be used to evaluate economic risks from climate change, or in the development of policy to attenuate damages.
\end{abstract}

\begin{fmtext}
\end{fmtext}
\maketitle

\section{Introduction}

As a complex global issue, the analysis of climate change requires input from numerous intellectual fields. Some economists have contributed by developing what they describe as ``Integrated Assessment Models'' (IAMs). IAMs contain components that map economic output to CO$_2$ generation, CO$_2$ generation to the increase in global average equilibrium temperature, and the increase in global average equilibrium temperature to the reduction in economic output, via what they call ``damage functions''.

In this review, we critically evaluate the empirical research done by economists to estimate these damage functions. We anchor the discussion around the Dynamic Integrated Climate Economy (DICE) model of W. D. Nordhaus \cite{nordhaussztorc2013}, because it was the first model of its kind, and it has been highly influential in determining climate change policy \cite{interagency2016,interagency2021}. We also examine economic damage estimates cited in the Intergovernmental Panel on Climate Change (IPCC) 5th Assessment Report (AR5) \cite{arent2014key}, and subsequent estimates \cite{burke2015globalnonlinear,kahn2019longterm}.

We start in Section \ref{conceptual} with conceptual issues including a serious misrepresentation of the scientific literature on climate tipping points and a corresponding trivialization of damages that arises from this misrepresentation. Then in Section \ref{methodological} we describe a series of methodological flaws behind these trivial damage estimates, including a discussion of how these obviously flawed methods became accepted and widely used by economists. In Section \ref{DICE_section}, we present a self-contained description of the DICE model and the role that climate damages play in it, including a discussion of its sensitivity with respect to more severe damages. Section \ref{Conclusion_section} concludes with our assessment of the 
contribution of economists to the climate debate and a plea that science-based approaches should take precedence.

\section{Conceptual Issues} 
\label{conceptual}

\subsection{Misrepresenting the scientific literature}
\label{tipping}

The existence of tipping points in the Earth's climate is now an established part of the scientific literature on climate change \cite{lenton2008tipping,cailenton2016risk,steffen2018trajectories,lenton2019climate,xu2020future}. The first paper to catalog and attempt to calibrate major tipping points that could be triggered by global warming reported on a survey of experts, who considered whether there were large-scale elements of Earth's climate system that could be triggered into abrupt and/or irreversible qualitative changes, causing significant qualitative changes to the climate \cite{lenton2008tipping}. These systems were selected  subject to the conditions that they must be ``subsystems of the Earth system that are at least subcontinental in scale'', and that they could be ``switched---under certain circumstances---into a qualitatively different state by small perturbations''. The researchers excluded consideration of ``systems in which any threshold appears inaccessible this century'' \cite[pp. 1786--87]{lenton2008tipping}.

Nine such systems were identified, all but one of which---the Indian Monsoon---could be tipped by temperature increases alone. The study identified two definite candidates for a tipping point this century that could be triggered by a rise of between 0.5$^\circ$C and 2$^\circ$C---``Arctic sea-ice and the Greenland ice sheet''---and noted that ``at least five other elements could surprise us by exhibiting a nearby tipping point'' \cite[p. 1792]{lenton2008tipping} before 2100. A key aspect of subsequent work 
\cite{steffen2018trajectories,cailenton2016risk,kriegler2009imprecise} has been the concept of ``tipping cascades'', whereby passing a threshold for one system---say, a temperature above which the Greenland ice sheet irreversibly shrinks---triggers causal interactions that increase the likelihood that other tipping elements undergo qualitative transitions---in this example, freshwater input to the North Atlantic increases the risk of a collapse of the Atlantic Meridional Overturning Circulation (AMOC---also referred to as the 'thermohaline circulation'). Such causal interactions can also be mediated by global temperature changes whereby tipping one system---e.g. the loss of Arctic summer sea-ice---amplifies global warming, increasing the likelihood that other other elements undergo a qualitative transition \cite{steffen2018trajectories}.

The paper \cite{lenton2008tipping} concluded that:

 \begin{quote}
Society may be lulled into a false sense of security by smooth projections of global change. Our synthesis of present knowledge suggests that a variety of tipping elements could reach their critical point within this century under anthropogenic climate change. The greatest threats are tipping the Arctic sea-ice and the Greenland ice sheet, and at least five other elements could surprise us by exhibiting a nearby tipping point. \cite[p. 1792]{lenton2008tipping}
\end{quote}

In contrast, the empirical estimates of economic damages from climate change reviewed below typically assume that climatic tipping points will not be triggered before 2100. One might surmise that this is because economists have not read this scientific literature, but that is not the case. In fact, Nordhaus \cite{nordhaus2013climate} cited \cite{lenton2008tipping} to justify \emph{not} including tipping points in economic damage functions:

\begin{quote}
There have been a few systematic surveys of tipping points in earth systems. A particularly interesting one by Lenton and colleagues examined the important tipping elements and assessed their timing\ldots \emph{The most important tipping points, in their view, have a threshold temperature tipping value of 3$^\circ$C or higher} (such as the destruction of the Amazon rain forest) \emph{or have a time scale of at least 300 years} (the Greenland Ice Sheet and the West Antarctic Ice Sheet). \emph{Their review finds no critical tipping elements with a time horizon less than 300 years until global temperatures have increased by at least 3$^\circ$C.} \cite[p. 60. Emphasis added]{nordhaus2013climate}

\end{quote}

Nordhaus also referenced \cite{lenton2008tipping} in the manual for DICE \cite{nordhaussztorc2013} to justify the use of a simple quadratic function to relate the increase in global temperature $\Delta T$ to the damages to global economic production $D(\Delta T)$:

\begin{quote}
The current version assumes that damages are a quadratic function of temperature change and does not include sharp thresholds or tipping points, \emph{but this is consistent with the survey by Lenton et al. (2008)}. \cite[p. 11. Emphasis added]{nordhaussztorc2013}
\end{quote}

The climate damage function is defined as the fraction by which future GDP would be reduced, relative to what it would be in the complete absence of climate change. In the current version of DICE, it is quantified as
\begin{equation}
    D(\Delta T)= a \times \Delta T^2, 
\label{quadratic_function}    
\end{equation}
where $a = 0.00227$. Nordhaus observed that this formula predicts ``damage of 2.1 percent of income at 3$^\circ$C, and 7.9 percent of global income at a global temperature rise of 6$^\circ$C'' \cite[p. 345]{nordhaus2018projections}. Table \ref{table:nordhausvlenton} contrasts the actual conclusions reached by Lenton et al. with Nordhaus's claims about those conclusions in \cite{nordhaus2013climate} and \cite{nordhaussztorc2013}.

\vspace{0.5 cm}

\begin{table}

    \centering 
    \caption{ Lenton et al.'s conclusions and their contradictory rendition by Nordhaus}
    \vspace{4mm} 
\label{table:nordhausvlenton}
\begin{tabular}{|p{6cm}|p{6cm}|}

\hline

Lenton 2008 ``Tipping elements in the Earth’s climate system'' \cite{lenton2008tipping} & Nordhaus 2013 DICE Manual \& \emph{The Climate Casino} \cite{nordhaussztorc2013,nordhaus2013climate} \\
\hline
{Society may be lulled into a false sense of security by smooth projections of global change. \cite[p.1792]{lenton2008tipping}} &  {The current version assumes that damages are a quadratic function of temperature change and does not include sharp thresholds or tipping points, but this is consistent with the survey by Lenton et al. (2008) \cite[p. 11]{nordhaussztorc2013}} \\
\hline
Our synthesis of present knowledge suggests that a variety of tipping elements could reach their critical point within this century under anthropogenic climate change.\cite[p.1792]{lenton2008tipping} & Their review finds no critical tipping elements with a time horizon less than 300 years until global temperatures have increased by at least 3$^\circ$C. \cite[p. 60]{nordhaus2013climate}\\
\hline

The greatest threats are tipping the Arctic sea-ice [$0.5$--2$^\circ$C warming] and the Greenland ice sheet [$1$--2$^\circ$C warming], and at least five other elements could surprise us by exhibiting a nearby tipping point. \cite[p.1792 \& Table 1, p. 1788]{lenton2008tipping}  & The most important tipping points, in their view, have a threshold temperature tipping value of 3$^\circ$C or higher (such as the destruction of the Amazon rain forest) \ldots \cite[p. 60]{nordhaus2013climate}\\

\hline

Arctic \ldots summer ice-loss threshold, \emph{if not already passed}, may be very close and a transition could occur well within this century. \cite[p.1789. Emphasis added]{lenton2008tipping} & \ldots or have a time scale of at least 300 years (the Greenland Ice Sheet and the West Antarctic Ice Sheet). \cite[p. 60]{nordhaus2013climate}\\

\hline

\end{tabular}
\end{table}

As Lenton later observed:
\begin{quote}
There is currently a huge gulf between natural scientists' understanding of climate tipping points and economists' representations of climate catastrophes in integrated assessment models (IAMs). \cite[p. 585]{lenton2013integrating}
\end{quote}

Nordhaus noted \cite[Note 11 to Chapter 5, p. 334)]{nordhaus2013climate} that his assertion that there are ``no \emph{critical} tipping elements with a time horizon less than 300 years until global temperatures have increased by at least 3$^\circ$C'' \cite[p. 60. Emphasis added]{nordhaus2013climate} relied in part on a table by Lenton in another publication, where Lenton constructed what he described as a ``simple ‘straw man’ example of tipping element risk assessment'' \cite[Table 7.1, p. 185]{richardson2011climate}. This table assessed the 8 tipping elements on two metrics, ``Likelihood of passing a tipping point (by 2100)'' and ``Relative impact of change in state (by 3000)'', and gave Arctic summer sea-ice the highest ranking on the first metric, and the lowest on the second. It is conceivable that this was the basis of Nordhaus's conclusion that there were no \emph{critical} tipping elements this century. However, Lenton explicitly noted that his impact rating was a \emph{relative} rating of the eight tipping elements against each other, not an absolute ranking of their climatic significance:

\begin{quote}
Impacts are considered in relative terms based on an initial subjective judgment (\emph{noting that most tipping-point impacts, if placed on an absolute scale compared to other climate eventualities, would be high}) \cite[p. 186. Emphasis added]{richardson2011climate}.
\end{quote}

The ``transition timescale'' column in Lenton's Table 1 \cite[p. 1788]{lenton2008tipping} \footnote{Table 1 ``Policy-relevant potential future tipping elements in the climate system'' summarises 15 tipping elements, headed by the 9 that could tip before 2100.} was also an estimate of the time the complete transition would take from its initial tipping, not the years until a tipping event would be triggered, as Nordhaus implied. While Lenton et al. did give a timeframe of more than 300 years for the complete melting of the Greenland Ice Sheet (GIS), for example, they noted that they considered only tipping elements whose fate would be decided this century:
\begin{quote}
Thus, we focus on the consequences of decisions enacted within this century that trigger a qualitative change within this millennium, and we exclude tipping elements whose fate is decided after 2100. \cite[p. 1787]{lenton2008tipping}
\end{quote}

Therefore, contrary to Nordhaus's assertions, Lenton's research warned that tipping elements were likely to be triggered by temperature rises expected this century, and noted that if triggered, these would have significant impacts upon the climate and biosphere, including its suitability for human life---and therefore, significant impacts upon the economy. Subsequent research by Lenton and associates has explicitly criticised the treatment of tipping points by economists in general (and Nordhaus in particular) \cite{lenton2013integrating}, calculated that including tipping points in Nordhaus's own DICE model can increase the ``Social Cost of Carbon'' (by which optimal carbon pricing is calculated) by a factor of greater than eight \cite{cailenton2016risk}, and proposed 2$^\circ$C as a critical level past which ``tipping cascades'' could occur \cite{lenton2013integrating,lenton2019climate,steffen2018trajectories}. 

This criticism of economists by scientists has not caused economists in general to recognise tipping points. While some economists have been influenced by Lenton's work on tipping points \cite{cailenton2016risk,lemoine2016economics}, subsequent use of the DICE model to inform policy (e.g. setting of the US Federal social cost of carbon \cite{interagency2016,interagency2021}) either ignored tipping points, assigned them such a low probability as to be irrelevant to damages, or introduced arbitrary discontinuities into damage functions at temperatures increases well above the 2$^\circ$C level at which the scientific literature indicates that tipping points become likely \cite{lenton2019climate,steffen2018trajectories}.

One instance of this is Nordhaus's treatment in DICE of the possibility that the AMOC might pass a tipping point \cite{nordhaus2000warming}. While he considers that it could cause a 25\% reduction of global GDP (later increased to 30\%), he assigns it probabilities of only 0.6\% for a 3$^\circ$C warming scenario in 2090 and 3.4\% for a 6$^\circ$C warming scenario in 2175. Nonetheless, given Nordhaus's other assumptions in DICE (a rate of risk aversion of 4 and an income elasticity of 0.1), the ``willingness to pay'' metric to avoid the AMOC catastrophe is 1\% of global GDP for the 3$^\circ$C scenario and 7\% for the 6$^\circ$C scenario. Thus, despite the very low assigned probabilities, this catastrophic risk comprises about half of total damages estimated by Nordhaus at 3$^\circ$C and the majority of damages at 6$^\circ$C.\footnote{Some economists specialising in climate change have argued that the shutdown of AMOC would on balance be economically \emph{beneficial}: ``The integrated assessment model FUND and a meta-analysis of climate impacts are used to evaluate the change in human welfare associated with a slowdown of the thermohaline circulation. We find modest, but by and large, positive effects on human welfare'' \cite[p. 602]{anthoffEstradaTol2016shutting}. For scientific assessments of what the shutdown of AMOC would entail, see \cite{boers2021observation,anthoffEstradaTol2016shutting}.} This indicates that, were this and other tipping points assigned more realistic probabilities, estimated damages would be much higher. Sure enough, inclusion of tipping point likelihoods in DICE from the scientific expert elicitation of \cite{kriegler2009imprecise} leads to much higher damages \cite{cailenton2016risk}.

We now turn to the numerical estimates that economists have made of the economic damages from climate change.

\subsection{Trivializing expected damages}

The  2014 IPCC Report \emph{Climate Change 2014: Impacts, Adaptation, and Vulnerability Part A: Global and Sectoral Aspects} \cite{field2014ipcc} included a chapter on the economic consequence of climate change, entitled ``Key economic sectors and services'' \cite{arent2014key}. The chapter included a scatter plot with 19 point estimates of temperature increase and global income change \cite[Figure 10.1, p. 690]{arent2014key}. These estimates ranged from (+1$^\circ$C, +2.3\%)---that is, a 2.3\% \emph{increase} in global income for a 1$^\circ$C increase in global temperature over pre-industrial levels---to (+5.4$^\circ$C, -6.1\%) on the temperature axis, and (+3.2$^\circ$C, -12.4\%) on the income axis, as shown in Figure \ref{fig:IPCC2014} below.\footnote{One study, \cite{mendelsohn2000country}, with two methodologies, ``Enumeration'' and ``Statistical'', was cited twice in these estimates. Figure \ref{fig:IPCC2014} includes damage estimates from \cite{burke2015globalnonlinear,kahn2019longterm}, which were published after \cite{arent2014key,field2014ipcc}}

\begin{figure}[h]
\centering
\caption{Adapted from \cite[Figure 10.1, p. 690]{arent2014key}: Estimates of the total impact of climate change for different values of temperature increase, and damage function from  DICE 2016R with 2018 damage coefficient of 0.00227 \cite[p. 345]{nordhaus2018projections}}
\includegraphics[width=0.8\textwidth]{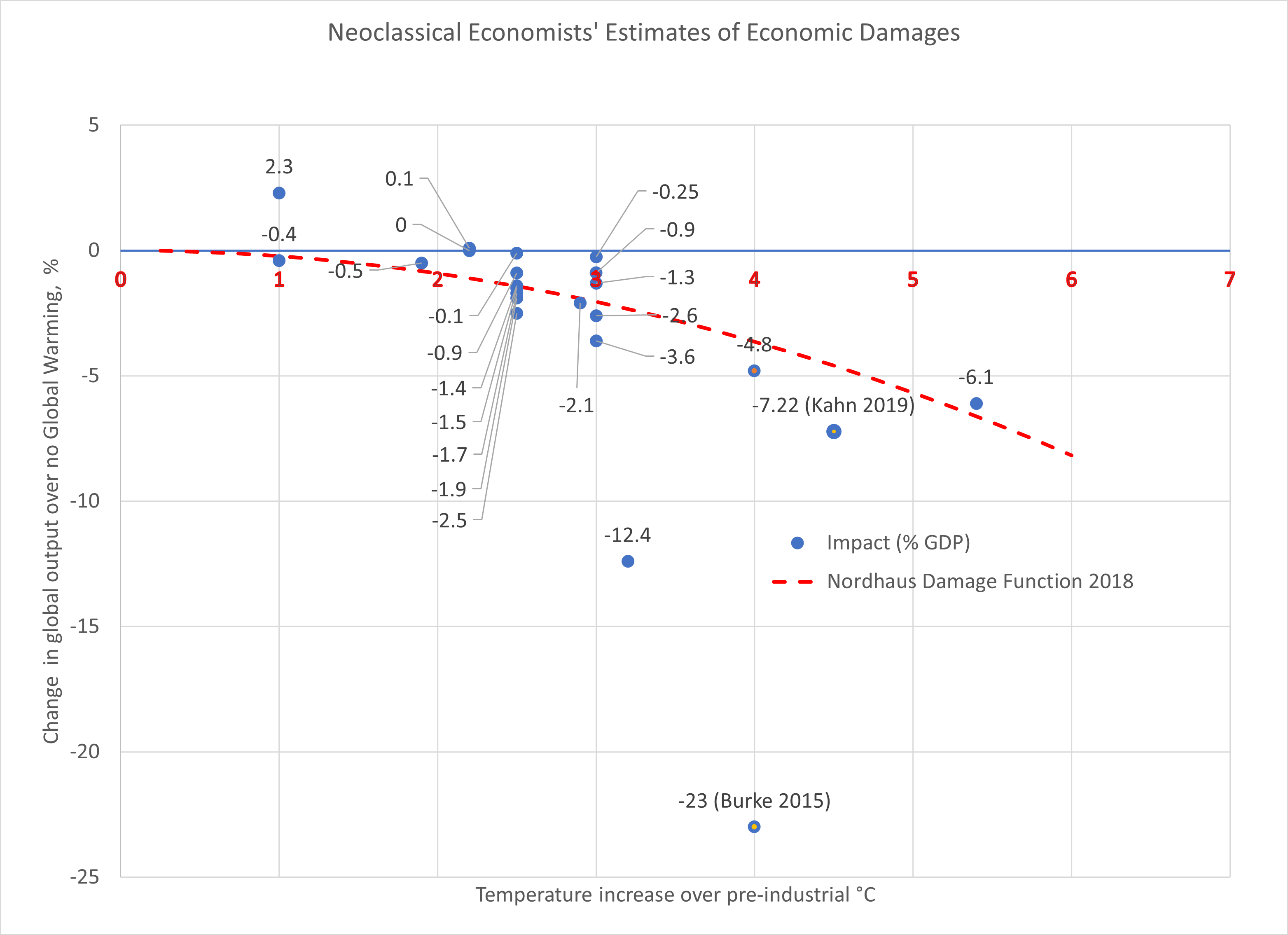}
\label{fig:IPCC2014}
\end{figure}

These estimates were in comparison to what annual global income would be in the early $22^{nd}$ century in the complete absence of global warming, and the chapter noted that ``Estimates agree on the size of the impact (small relative to economic growth)'' \cite[p. 690]{arent2014key}. This is an assertion that the global economy would only be mildly impacted by global warming---even in the event of temperature increasing by 5.4$^\circ$C over pre-industrial levels by as soon as 2100 \cite{roson2012climate}. As the Executive Summary to that chapter put it:

\begin{quote}
For most economic sectors, the impact of climate change will be small relative to the impacts of other drivers (\emph{medium evidence, high agreement}). Changes in population, age, income, technology, relative prices, lifestyle, regulation, governance, and many other aspects of socioeconomic development will have an impact on the supply and demand of economic goods and services that is large relative to the impact of climate change. \cite[p. 662]{arent2014key}
\end{quote}

\begin{figure}[h!]
	\centering
	\caption{Nordhaus's downward revisions to his damage function.}
\includegraphics[width=0.8\textwidth]{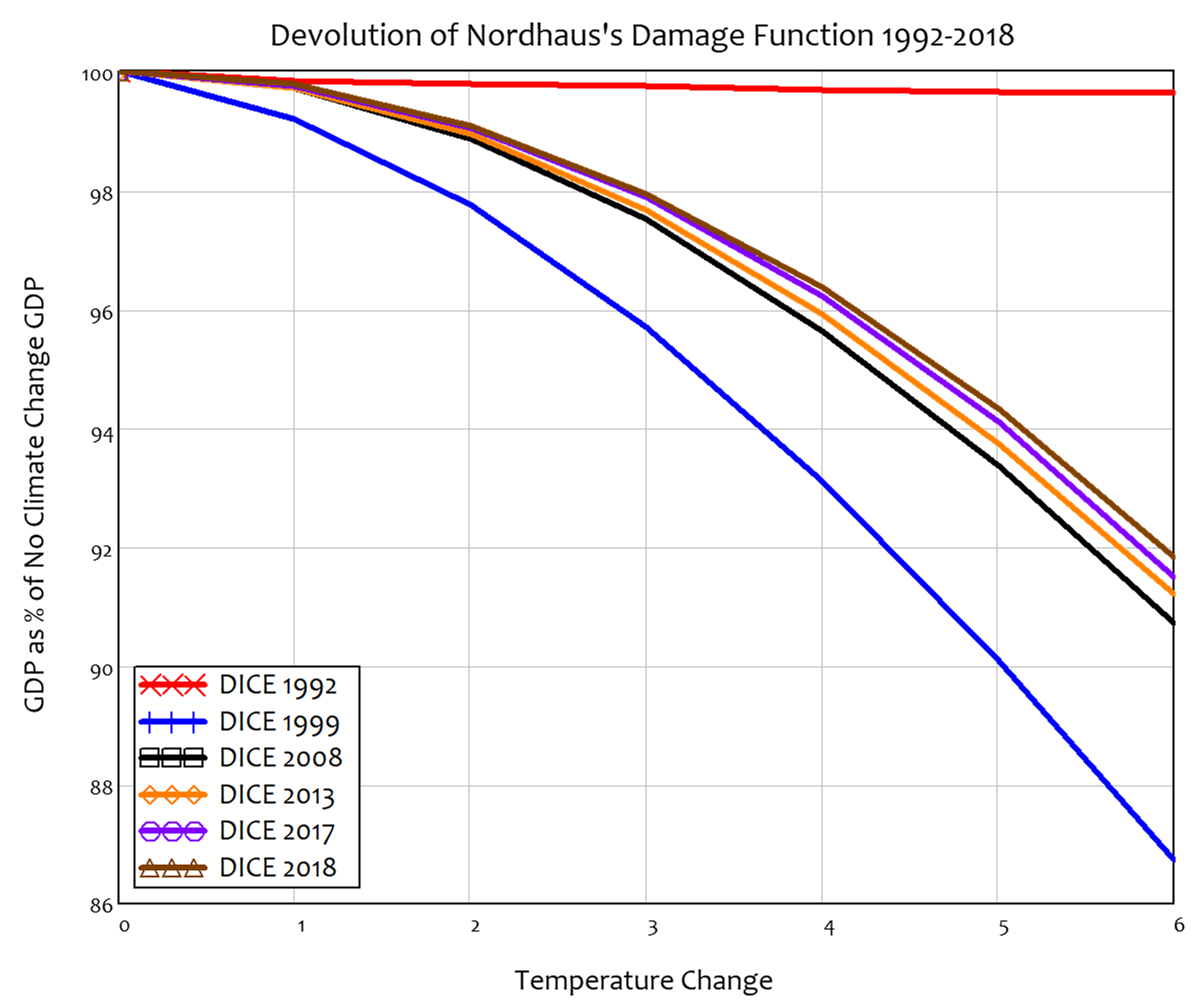}
\label{fig:nordhausdamagenumbers}
\end{figure}

Figure \ref{fig:IPCC2014} also plots the damage function used by Nordhaus in 2018 \cite{nordhaus2018projections}, which was calibrated against a subset of this data (see \cite{nordhausmoffat2017}). As shown in Table \ref{tab:nordhausdamagenumbers} and Figure \ref{fig:nordhausdamagenumbers}, Nordhaus has used different functional forms and generally tended to adopt \emph{less} convex functions with each revision of the model.

\begin{table}[htb]
\caption{Nordhaus damage function specifications}
    \centering 
    \vspace{4mm} 
\begin{tabular}{p{2cm}p{4cm}p{5cm}}

\bf Year& \bf Function $1-D(\Delta T)$ & \bf Parameters\\
1992 & $\frac{1}{1+\frac{a}{9}+\sqrt \Delta T}$ \cite[p. 85]{nordhaussztorc2013} & $a=0.0133$ \cite[p. 83]{nordhaussztorc2013} \\
1999 & $\frac{1}{1+b\times \Delta T+a\times \Delta T^2}$ \cite[p. 88]{nordhaussztorc2013}  & $b=0.0045 , a=0.0035$ \cite[p. 86]{nordhaussztorc2013} \\
2008   & $\frac{1}{1+a\times \Delta T^2}$ \cite[p. 94]{nordhaussztorc2013} & $a=0.0028388$ \cite[p. 91]{nordhaussztorc2013}\\
2013  & $\frac{1}{1+a\times \Delta T^2}$ \cite[p. 100]{nordhaussztorc2013}  & $a=0.00267$ \cite[p. 97]{nordhaussztorc2013} \\
2017 & $1-a\times \Delta T^2$ \cite[p.1]{nordhaus2017revisiting} & $a=0.00236$ \cite[p.1]{nordhaus2017revisiting} \\
2018 & $1-a\times \Delta T^2$ & $a=0.00227$ \cite[p. 345]{nordhaus2018projections} 

\end{tabular}
\label{tab:nordhausdamagenumbers} 
\end{table}

Figure \ref{fig:nordhausdamagenumbers} plots these damage functions against temperature increase: with the sole exception of the change from the initial form in 1992 to a quadratic-based formula, the impact of each change has been to reduce the predicted damages from climate change.

As mentioned in the previous section, given the scope for fundamental changes to the climate arising from several tipping points in the Earth's climate, it is difficult to reconcile the sanguine conclusion from the estimates in Figure \ref{fig:IPCC2014}, and Nordhaus's damage function, with the level of damage to the biosphere expected by climate scientists \cite{lenton2013integrating,cailenton2016risk,steffen2018trajectories,lenton2019climate,xu2020future,kriegler2009imprecise,sherwood2010adaptability,schellnhuber2006avoiding}.

For simple examples suggesting that economic damages should be much higher, consider the following. A critical feature of human physiology is our ability to dissipate internal heat by perspiration. To do so, the external air needs to be colder than our ideal body temperature of about 37$^\circ$C, and dry enough to absorb our perspiration as well. This becomes impossible when the combination of heat and humidity, known as the ``wet bulb temperature'', exceeds 35$^\circ$C. Above this level, we are unable to dissipate the heat generated by our bodies, and the accumulated heat will kill a healthy individual within a few hours. Scientists have estimated that a 3.8$^\circ$C increase in the global average temperature would make Jakarta's temperature and humidity combination permanently fatal for humans, while a 5.5$^\circ$C increase would mean that even New York would experience 55 days per year when the combination of temperature and humidity would be deadly \cite[Figure 4, p. 504]{mora2017global}.

Temperature also affects the viable range for all biological organisms on the planet. Scientists have estimated that a 4.5$^\circ$C increase in global temperatures would reduce the area of the planet in which life could exist by 40\% or more, with the decline in the livable area of the planet ranging from a minimum of 30\% for mammals to a maximum of 80\% for insects \cite[Figure 1, p. 792]{warren2018projected}.

It would therefore be tempting for climate scientists to simply outright dismiss the estimates of damages in Figure \ref{fig:IPCC2014}. Nonetheless, since politicians pay disproportionate attention to the opinions of economists in formulating most government policies \cite{hirschmanberman2014economists,pusey1991economic,decanio2003economic,interagency2016,interagency2021},\footnote{``it is undeniably the case that economic arguments, claims, and calculations have been the dominant influence on the public political debate on climate policy in the United States and around the world\ldots It is an open question whether the economic arguments were the cause or only an ex-post justification of the  decisions made\ldots but there is no doubt that economists have claimed that their calculations should dictate the proper course of action.'' \cite[p. 4]{decanio2003economic}} this trivialisation of the dangers of climate change by economists has had serious deleterious effects on the human response to climate change.

Accordingly, we argue that it is of upmost importance that climate scientists critically engage with economists on this issue. To this end, in the next section we describe what we consider to be flawed methods used by economists to arrive at the estimates mentioned above.

\section{Methodological Issues}
\label{methodological} 

\subsection{Equating weather and climate}

The first numerical estimate made by economists of the impact of climate change on economic output was generated by Nordhaus in 1991, in the same paper that introduced the cost-benefit framework for the analysis of climate change \cite{nordhaus1991slow}. He predicted a 0.25\% fall in income for a 3$^\circ$C increase in global average temperature, while making qualifications to this estimate---which he described as ``ad hoc''---to suggest damages may be 1\% of GDP, but were unlikely to be more than 2\% of GDP:

\begin{quote}
damage from a (3$^\circ$C) warming is likely to be around $\frac{1}{4}$\% of  national income for United States in terms of those variables we have been able to quantify. This figure is clearly incomplete, for it neglects a number of areas that are either inadequately studied or inherently unquantifiable. We might raise the number to around 1\% of total global income to allow for these unmeasured and unquantifiable factors, although such an adjustment is purely ad hoc. It is not possible to give precise error bounds around this figure, but my hunch is that the overall impact upon human activity is unlikely to be larger than 2\% of total output\ldots \cite[p. 933]{nordhaus1991slow}
\end{quote}

This paper was not used in the IPCC 2014 Report.\footnote {Given the history of how these estimates were assembled, this is more likely to be an oversight than any application of quality control. Richard Tol, one of the two Coordinating Lead Authors of this chapter, has a history of data errors in his published papers: two revisions were published to his 2009 paper ``The Economic Effects of Climate Change'' \cite{tol2009economic}, due to errors that included omitting negative signs from damage estimates: see \cite{tolerrors2015}. This saga is discussed in detail in a post in \href{https://statmodeling.stat.columbia.edu/2014/05/27/whole-fleet-gremlins-looking-carefully-richard-tols-twice-corrected-paper-economic-effects-climate-change/}{Statistical Modeling, Causal Inference and Social Science} and 2 subsequent posts. Nordhaus and Moffat noted further uncorrected errors in 2017 \cite{nordhausmoffat2017}.} However, its method of estimating damages from climate change---later described as ``Enumeration'' by Tol \cite{tol2009economic}---was replicated by 11 of the 18 studies \cite{nordhaus2008question,bosello2012assessingwp,bosello2012assessing,hope2006marginal,mendelsohn2000country,tol2002estimates,nordhaus2000warming,plambeck1996page95,nordhaus1996regional,fankhauser1995valuing,nordhaus1994managing} cited by the IPCC Report \cite{arent2014key}. A key element of this method was the assumption, \emph{made without any explanation beyond what is quoted below}, that only activities directly exposed to the weather would be affected by climate change:

\begin{quote}
Table 5 shows a sectoral breakdown of United States national income, where the economy is subdivided by the sectoral sensitivity to greenhouse warming. The most sensitive sectors are likely to be those, such as agriculture and forestry, in which output depends in a significant way upon climatic variables. At the other extreme are activities, such as cardiovascular surgery or microprocessor fabrication in 'clean rooms', which are undertaken in carefully controlled environments that will not be directly affected by climate change.  Our estimate is that approximately 3\% of United States national output is produced in highly sensitive sectors, another 10\% in moderately sensitive sectors, and \emph{about 87\% in sectors that are negligibly affected by climate change}. \cite[p. 930. Emphasis added]{nordhaus1991slow}
\end{quote}

As this quote indicates, this assumption was used to omit entire sectors of the economy from consideration---including manufacturing, which represented 26\% of US GDP in 1991, retail and wholesale trade (28\%), and government (14\%)---see Table \ref{table:nordhaus1991breakdown}.

\begin{table}[htb]

    \centering 
    \caption{Extract from Nordhaus's breakdown of economic activity by vulnerability to climatic change in the US \cite[Table 5, p. 931]{nordhaus1991slow}}
    \vspace{4mm} 
\label{table:nordhaus1991breakdown}
\begin{tabular}{|p{7cm}|p{4cm}|}
\hline
{\bf ``Negligible Effect'' Sectors} & {\bf Percentage of total economy}\\
\hline
	& \\
\hline
Manufacturing and mining &	26.0\\
\hline
Other transportation and communication	& 5.5\\
\hline
Finance, insurance, and balance real estate	& 11.4\\
\hline
Trade and other services	& 27.9\\
\hline
Government services & 14.0\\
\hline
Rest of world & 2.1\\
\hline
{\bf Total ``negligible effect''	} & {\bf 86.9}\\
\hline

\end{tabular}
\end{table}

The only way to make sense of Nordhaus's assumption is that he equated ``affected by climate change'' to ``affected by the weather'', since the only unifying feature of the activities he assumed would be unaffected is that they occur either indoors or underground. Since these sectors, representing 87\% of the economy, are not directly affected by the weather, he expressed an inability to understand how they could be affected by climate change:

\begin{quote}
for the bulk of the economy---manufacturing, mining, utilities, finance, trade, and most service industries---it is difficult to find major direct impacts of the projected climate changes over the next 50 to 75 years. \cite[p. 932]{nordhaus1991slow}
\end{quote}

The same presumption, that only activities that are directly exposed to the weather will be affected by climate change, was repeated by the economics chapter of the 2014 IPCC Report as a ``Frequently Asked Question'':
\begin{quote}
{\bf FAQ 10.3 | Are other economic sectors vulnerable to climate change too?} Economic activities such as agriculture, forestry, fisheries, and mining are exposed to the weather and thus vulnerable to climate change. Other economic activities, such as manufacturing and services, \emph{largely take place in controlled environments and are not really exposed to climate change}. \cite[p. 688. Emphasis added]{arent2014key}
\end{quote}

The only change between Nordhaus in 1991 \cite{nordhaus1991slow} and the IPCC Report in 2014 \cite{arent2014key} was that the IPCC noted that mining was ``exposed to the weather''---presumably in recognition of open-cut mining. However, none of the studies actually considered the impact of the weather on mining, while the industries surveyed in \cite{nordhaus2008question,bosello2012assessingwp,bosello2012assessing,hope2006marginal,mendelsohn2000country,tol2002estimates,nordhaus2000warming,plambeck1996page95,nordhaus1996regional,fankhauser1995valuing,nordhaus1994managing} exclude the same list of industries excluded from consideration by Nordhaus---see the ``Coverage'' column in Table \ref{table:ipccstudies}. Similarly, the IPCC special report in 2018 on Global Warming of 1.5$^\circ$C \cite{masson2018} section 3.4.9 ``Key Economic Sectors and Services'', lists tourism, energy (focusing on change in energy demand for heating and cooling), and transportation (focusing primarily on reduced shipping costs thanks to an ice-free Arctic) as economic sectors affected by global warming, but makes no mention of impacts on manufacturing or services.

This assumption that only economic activities that are exposed to the weather will be affected by climate change can be rejected on at least three grounds.

Firstly, some manifestations of climate change---such as increased wildfires\footnote{For some recent news coverage, see for example \url{https://apnews.com/8e4e0818146a72c713de625e902f9962}.} \cite{dowdy2019future} and floods \cite{kulp2019new}---will affect factories and offices at least as much as they affect the output from those factories and offices.

Secondly, indoor activities will be unable to continue if temperature increases of the scale contemplated by economists, such as the 6$^\circ$C increase that Nordhaus asserted would reduce GDP by only 7.9\% \cite[p. 345]{nordhaus2018projections}, mean that the places in which factories and offices are located become uninhabitable---see studies such as \cite{mora2017global}, which find that lethal temperature levels could affect areas currently home to 74\% of the world's population by 2100 \cite[p. 501]{mora2017global}. Factories without workers produce zero output.

Thirdly, climate change---including impacts on biodiversity and resource depletion as well as global warming---will affect the availability and costs of essential inputs to production from the environment: manufacturing cannot occur without non-manmade inputs from the environment, most obviously energy \cite{keen2019note}, but clearly also agricultural and mineral inputs. Climate-change induced stress on power grids and agriculture will clearly impact manufacturing and services \cite{schewe2019state}. We speculate that Nordhaus may have ignored this obvious point because the Cobb-Douglas production function (see equation \eqref{CD_production} below) used in his DICE model (and most Neoclassical economic models today), ignores environmental inputs to production, and implies that production can occur with inputs of labour and machinery alone \cite{keen2019note}.

\subsection{Using spatial variations in the existing climate as a proxy for climate change}

The second most frequent method used to generate the numbers in Figure \ref{fig:IPCC2014}, labelled as the ``statistical'' method \cite{maddison2011impact,nordhaus2006geography,rehdanz2005climate,maddison2003amenity,mendelsohn2000country} by the IPCC, was described as follows by one of the two co-authors of the economics chapter of the IPCC 2014 Report \cite[p. 659]{arent2014key}, Richard Tol:

\begin{quote}
It is based on direct estimates of the welfare impacts, using observed variations (across space within a single country) in prices and expenditures to discern the effect of climate. \emph{Mendelsohn assumes that the observed variation of economic activity with climate over space holds over time as well}; and uses climate models to estimate the future effect of climate change. Mendelsohn's estimates are done per sector for selected countries, extrapolated to other countries, and then added up. \cite[p. 32. Emphasis added]{tol2009economic}
\end{quote}

\begin{figure}
\centering
\caption{Data and quadratic fit on Temperature and Gross State Product per capita}
\includegraphics[width=0.9\textwidth]{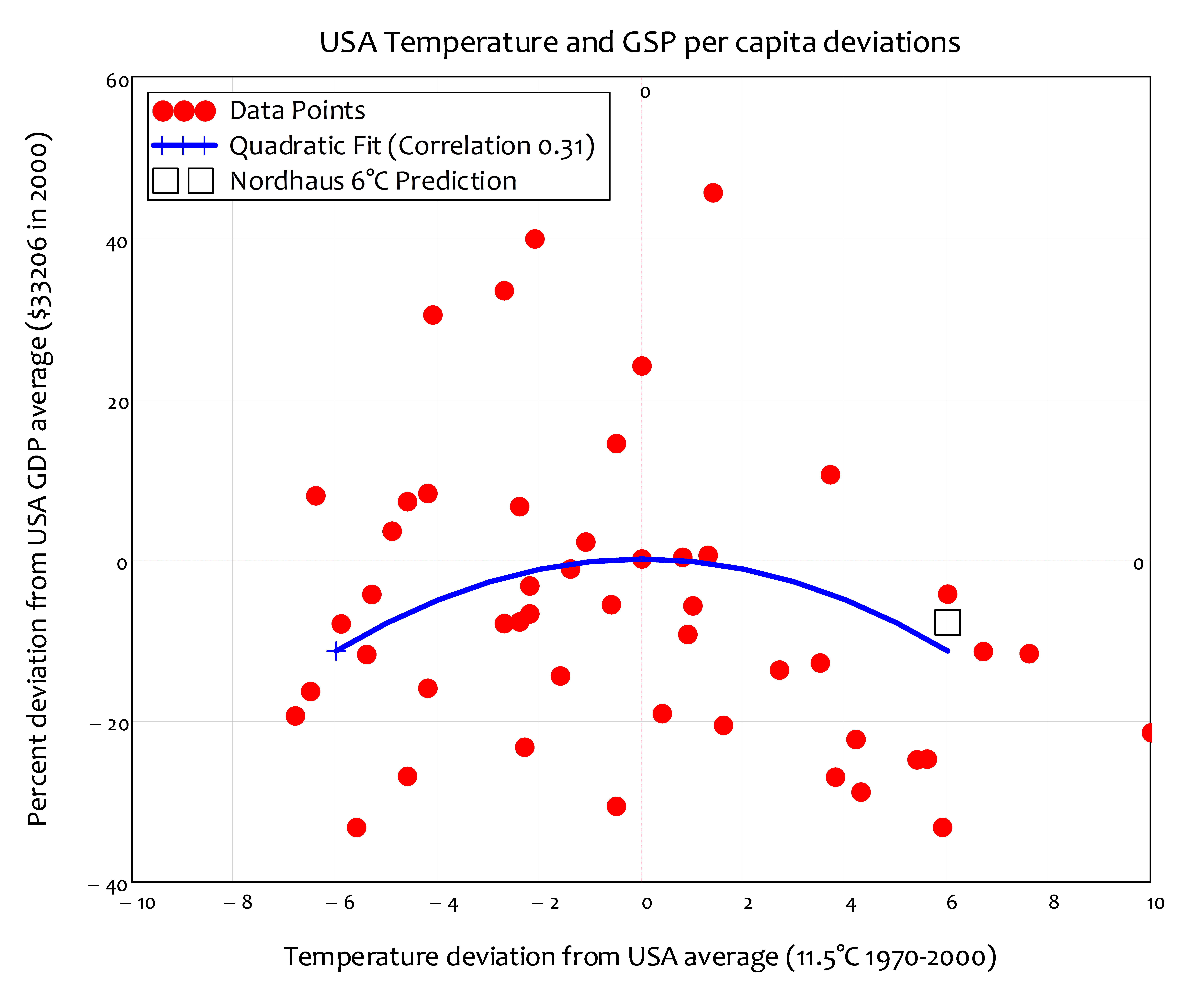}
\label{fig:f3gdpvtemp}
\end{figure}

This method, which is illustrated by Figure \ref{fig:f3gdpvtemp}, derives a statistical fit between temperature today and income today in the USA,\footnote{Later papers have developed global temperature and GDP comparisons, see for example \cite{nordhaus2006geography}.} typically using a simple regression on the coefficient for a quadratic function of the temperature deviation from the average national temperature. It then uses the parameter derived from this regression as the predictor of the impact of global warming on future GDP.

In the illustration shown in Figure \ref{fig:f3gdpvtemp}, the quadratic coefficient is -0.318\% of GDP per 1$^\circ$C squared rise in temperature over pre-industrial levels. While small, this coefficient is nonetheless \emph{larger} than that used by Nordhaus in 2018 to produce his prediction that a 6$^\circ$C increase in global average temperature will result in a decrease of global income by only 7.9\% relative to what it would have been otherwise in the absence of a temperature rise \cite[p. 345]{nordhaus2018projections} ---see equation \eqref{quadratic_function}.\footnote{The range of the quadratic in Figure \ref{fig:f3gdpvtemp} is restricted to the magnitude of temperature changes for which Nordhaus has given predictions of damages in refereed papers.}

 Assessing future climate damages by relating current temperature and income  is quite clearly problematic. Figure \ref{fig:f3gdpvtemp} focuses on present and regional variability within the United States alone, suggesting existence of an optimal climate from which deviations can be related to lower productivity. However, the fit is weak. Statistically, the fit to temperature explains just 10\% of Gross State Product (GSP) per capita variability, if we assume that all locations are statistically independent, and less if we allow for interactions between regions through inter-regional trade. The most reasonable conclusion is that there is effectively no significant relationship between contemporaneous GDP and temperature.
 
However, even the weak fit is statistically misplaced as a guide to future climate damages, because the underlying premise is flawed. The relationship between \emph{regional} temperature variations and \emph{regional} economic output \emph{at a given time} has, in principle, no bearing on \emph{global} increases {\em over time} in average surface temperature and \emph{global} economic activity. Effectively, what economists have assumed is that the climate-economy relation, for a specific region of Earth at a specific time, under the condition of a stable level of atmospheric CO$_2$, maps onto the climate-economy relation for the world as a whole as CO$_2$ levels rise.

This assumption commits two fundamental errors. First, any current relationship between GDP and temperature must reflect regional climate resilience that has developed over centuries. Obviously, the Gross World Product of the planet taken as a whole is an economically isolated system, since Earth is unable to engage in interplanetary trade. But the same quite clearly does not apply to regional economies. For example, Alaska and Maryland are two climate extremes in the United States that have similar GSP per capita, despite their large differences in average temperature.\footnote{The GSP data used in Figure \ref{fig:f3gdpvtemp} is sourced from the Bureau of Economic Analysis \hyperlink{https://apps.bea.gov/}{SAINC1} Tables. The figures for Alaska and  Maryland in 2020 were $\$64,780$ and $\$68,258$ respectively. Temperature data comes from \hyperlink{https://www.currentresults.com/Weather/US/average-annual-state-temperatures.php}{Current Results}. See Table \ref{tempgdpdata}.
} A prominent climate economist inferred on social media that:

\begin{quote}
    10K is less than the temperature distance between Alaska and Maryland (about equally rich), or between Iowa and Florida (about equally rich). Climate is not a primary driver of income.\footnote{See  https://twitter.com/RichardTol/status/1140591420144869381?s=20}
\end{quote}

Effectively, what is ignored in this argument is the importance of trade. Trade smooths regional disparities in a manner that is likely to change with the climate.  The USA has been agriculturally prosperous because of at least three historical coincidences: a currently hospitable climate suitable for growing crops; a past hospitable climate suitable for the establishment of topsoil; and the development of a single political, social, and economic system for the distribution of agricultural output across that system. Thus, current regional productivity, and its distribution, depends not only on current climate but also on centuries to millennia of prior climatic states, and centuries of political and economic evolution designed around regional exchange. 

Therefore, supposing a rapid change in global average temperature that shifts the climate suitable for growing crops further north, but at a speed that far exceeds the rate at which new topsoil is created, and in a manner that transgresses political and social boundaries adapted to agricultural productivity, the impact on GDP can be expected to be significantly greater than that reflected by the quadratic curve used in DICE and illustrated in Figure \ref{fig:f3gdpvtemp}. Trade may limit the sensitivity of production to local climate by offering access to economies in other climates, but not if those economies are also damaged by climate change.

Consider just one key commodity, grain. Neither Florida nor Alaska, lying at either end of the temperature extremes of the current climate of the USA, would be nearly as economically productive were it not for being able to access this primary human food source from Iowa. If, on the other hand, the climate of Iowa starts to approximate that of Florida, and grain production there fails, it will not be replaced at higher latitudes due to the poorer topsoil. Even with continued inter-state trade and new patterns of international trade, equivalently high levels of agricultural prosperity would be difficult to sustain \cite{xu2020future}.

The second, and more fundamental, concern is that the physics of regional climate variability differs from global climate variability, so the economic response should also differ. Climate change can be expected to introduce new climate extremes not currently experienced in any economically active region. For example, high wet-bulb temperatures that are known to be life-threatening to mammals (including humans), and that have not previously existed in the Holocene climate, are expected to develop over large, densely-inhabited areas. As Sherwood and Huber \cite{sherwood2010adaptability} note when citing two references in this economic literature, 

\begin{quote}
    The damages caused by  $10^\circ$C of warming are typically reckoned at 10–30\% of world GDP \cite{nordhaus2000warming,hope2006marginal}, roughly equivalent to a recession to economic conditions of roughly two decades earlier in time. While undesirable, this is hardly on par with a likely near-halving of habitable land, indicating that current assessments are underestimating the seriousness of climate change.  \cite[p. 9555]{sherwood2010adaptability}
\end{quote}

Further, through tipping elements, the qualitative characteristics of  global climate could change profoundly, including its spatial patterns and modes of temporal variability \cite{lenton2013integrating,Rodgers2021}. If the spatial-temporal pattern of climate variability changes with the mean state, this future challenges any drawing of inferences from the current spatial-temporal pattern. For example, a possible shutdown of the AMOC would have global impacts, including threats to monsoons in West Africa and India as well as to the equability of Western Europe's climate. In addition, sea level rise this century could expose hundreds of millions of people to flooding \cite{kulp2019new} particularly in the tropics \cite{hooijer2021globallidar}. None of these key aspects can be mathematically characterized by a simple fit (quadractic or otherwise) of current temperature to current income.

When global warming changes the mean climate state, some regional climates are eliminated while new climates are introduced, shifting the distribution of regional climates across the face of the Earth. Of course, not all climate shifts would be regionally detrimental \cite{mendelsohn2000country}, but overall a mismatch would develop between the new global climate and the conditions under which human populations developed \cite{xu2020future}. In a changing world, potentially billions will be presented with the option to stay put and try to adapt, or to migrate to more climatologically favorable regions \cite{xu2020future}. These geopolitical and economic challenges lie well beyond the narrow economic considerations demonstrated by the equilibrium quadratic fit of Figure \ref{fig:f3gdpvtemp}.

As a final remark about using spatial variations to estimate damages from climate change, we notice that the (3.2$^\circ$C,-12.4$\%$) point on Figure \ref{fig:IPCC2014}, taken from \cite{maddison2011impact}, appears to be a significant outlier from the other numbers. However, it was based on the same assumption that geographic data could serve as a proxy for climate change:

\begin{quote}
The approach that we will go on to describe in more detail deals exclusively with non-market impacts \emph{whilst making use of spatial variations in the existing climate as an analogue for climate change}.  \cite[p. 2438. Emphasis added]{maddison2011impact} 
\end{quote}

The fact that this appears to imply a possible ``tipping point'' of impacts at 3.2$^\circ$C is thus coincidental. \footnote{The same observation applies to the post-IPCC 2014 estimate by Burke \cite{burke2015globalnonlinear}, which we discuss in section  \ref{post-ipcc2014}.}

\subsection{Computable General Equilibrium Models}

Nordhaus's DICE model is based on the Neoclassical growth model first developed by Ramsey \cite{ramsey28mathematicaltheoryofsaving}, in which a single individual, treated as a ``representative agent'' for all of society, maximizes its utility over an infinite time horizon, by producing and consuming a single commodity (see Section \ref{DICE_section} below). This modelling approach dominates economics today, largely supplanting an earlier approach called ``Computable General Equilibrium'' (CGE), in which production of multiple commodities using commodities as inputs was modelled using input-output matrices. CGE models were the third most common technique to generate estimates of GDP reduction due to temperature change from global warming, with two references \cite{bosello2012assessing,roson2012climate} producing three estimates.

While CGE models have the strength of including the dependence of the production of commodities on myriad inputs, including other commodities \cite{sraffa1960production} and inputs from the natural environment, their reliance upon convergence to equilibrium prices and quantities ignores the instability of input-output matrices, which are unstable in either prices or quantities, given the ``dual stability theorem'' \cite[pp. 132-146]{blatt1983dynamic} (a consequence of the Perron-Frobenius theorem). See also \cite{decanio2003economic} for the weaknesses of disaggregated economic models as applied to climate change.

However, for the purposes of this review, the main weaknesses of the CGE approach are the characteristics it shares with the other methods criticized here. Both papers cited by the IPCC focus upon the same weather-exposed economic sectors (see the ``Coverage'' column for \cite{bosello2012assessing,roson2012climate}), and ignore the existence of tipping points.

The economic damage assessments from the General Equilibrium models cited here thus share the ``Enumerative approach'' assumption that only industries directly exposed to the weather will be affected by climate change. This can be seen in the list of phenomena and industries to which these studies have been applied: ``sea-level rise, agriculture, health, energy demand, tourism, forestry, fisheries, extreme events, energy supply\ldots'' \cite[p. 4]{bosello2012assessing}: the manufacturing, services, and government sectors, which represent the bulk of the economy, are conspicuously absent. On tipping points, Roson and van der Mensbrugghe note of their own study that ``Catastrophic events and extreme weather are not taken into account'' \cite[p. 274]{roson2012climate}.

The impact of the CGE methodology is also generally to reduce the already underestimated economic damages from climate change. The inputs to these models are the ``partial equilibrium'' estimates of individual damages -- the effect of higher temperatures on land use, agriculture, mortality, etc., as estimated by the other papers detailed in Table \ref{table:ipccstudies}. CGE methods then model an assumed adaptation to a welfare-maximizing equilibrium, which reduces the impact of these ``partial equilibrium'' inputs. Consequently, ``general equilibrium estimates tend to be lower, in absolute terms, than the bottom-up, partial equilibrium estimates. The difference is to be attributed to the effect of market-driven adaptation'' \cite[p. 20]{bosello2012assessing}.

\subsection{Survey of non-experts}

The numerical estimate in Figure \ref{fig:IPCC2014} and Table \ref{table:ipccstudies} of a 3.6\% fall in GDP for a $3^\circ$C increase in global average temperature by 2090, was derived from what the IPCC Report characterised as ``expert elicitation'' \cite{nordhaus1994expert}. This was a survey by Nordhaus of the opinions of 19 individuals---including Nordhaus himself---on the impact that increases in global average temperature would have on global economic output. Nordhaus described them as including ten economists, four ``other social scientists'', and five ``natural scientists and engineers'', including three climate scientists,\footnote {``Stephen Schneider (climatology). Professor of biological science, Stanford University\ldots Paul Waggoner (meteorology and agricultural science). Distinguished scientist at the Connecticut Agricultural Experiment Station\ldots Robert White (atmospheric science and engineering). President, National Academy of Engineering.'' \cite[p. 51]{nordhaus1994expert}} but also noted that eight of the economists came from ''other subdisciplines of economics (those whose principal concerns lie outside environmental economics)'' \cite[p. 48]{nordhaus1994expert}. Though they all had some connection to climate change,\footnote{The weakest connection was for Larry Summers: ``As chief economist at the World Bank, he supervised studies and the writing of World Development Report 1992, which surveyed development and the environment, and interacted with authors and the writing of background papers on the economic aspects of global warming''.  \cite[p. 51]{nordhaus1994expert}} as Nordhaus details \cite[p. 51]{nordhaus1994expert}, they were not experts in the sense of the more recent expert elicitation on tipping points \cite{lenton2008tipping,kriegler2009imprecise}. Nor were they ``encouraged to remain in their area of expertise''  \cite[p. 10]{lenton2008supplement}, as evidenced by the remarks on biodiversity made by one of the economists, which Nordhaus contrasted with the reservations expressed by a scientist:

\begin{quote}
One economist (4) stated there would be little impact through ecosystems: ''For my answer, the existence value [of species] is irrelevant---I don't care about ants except for drugs.'' By contrast, another respondent cautioned that the loss of genetic potential might lower the income of the tropical regions substantially. This difference of opinion is on the list of interesting research topics. \cite[p. 50]{nordhaus1994expert}
\end{quote}

Nordhaus's key question asked for a prediction of the effect on future annual GDP in three temperature scenarios: (A) a 3$^\circ$C rise by 2090, (B) a 6$^\circ$C rise by 2175, and (C) a 6$^\circ$C rise by 2090. The figure used by the IPCC from this paper was the mean for scenario A, of a 3.6\% fall in GDP for a 3$^\circ$C increase by 2090. But the most interesting features of this paper were the range of responses, and the disconnect between the opinions of economists and those of the scientists---only 3 of whom were scientists with a specialisation in climate change, and one of whom refused to answer this question. This scientist stated in his dissension that:
\begin{quote}
I marvel that economists are willing to make quantitative estimates of economic consequences of climate change where the only measures available are estimates of global surface average increases in temperature. As [one] who has spent his career worrying about the vagaries of the dynamics of the atmosphere, I marvel that they can translate a single global number, an extremely poor surrogate for a description of the climatic conditions, into quantitative estimates of impacts of global economic conditions. \cite[p. 51]{nordhaus1994expert}
\end{quote}

\begin{table}
    \centering 
    \caption{Summary statistics from Nordhaus's survey \cite{nordhaus1994expert}}
    \vspace{4mm} 
\begin{tabular}{|c|c|c|c|c|c|c|c|c|c|c|}

\hline

\multicolumn{3}{|c}{\makecell {\\ Scenario}} & \multicolumn{4}{|c} {\makecell {\\ Average estimates\\ (50th percentile)}} &\multicolumn{4}{|c|}{\makecell {Individual Percentile Ranges}}\\

\cline{8-11}

\multicolumn{3}{|c}{}& \multicolumn{4}{|c}{} & \multicolumn{2}{|l|}{\makecell{10th \\ (small impact)}}& \multicolumn{2}{l|}{\makecell {90th \\ (large impact)}}\\

\hline

Label	& \(\Delta T\)	&Date	&Min	&Max	&Median	&Mean	&Min&	Max&	Min	&Max\\
\hline
A	&3$^\circ$C	&2090&	0.0\%&	-21.0\%&	-1.9\%	&-3.6\%&	2.0\%&	-10.0\%&	-0.5\%&	-31.3\%\\
\hline
B&	6$^\circ$C	&2175&	0.0\%&	-35.0\%	&-4.7\%&	-6.1\%&	1.0\%&	-10.0\%	&-1.5\%&	-50.0\%\\
\hline
C	& 6$^\circ$C&	2090& 	-0.8\%&	-62.0\%&	-5.5\%&	-10.4\%&	-1.0\%&	-20.0\%&	-3\%&	-100.0\%\\
\hline

\end{tabular}
\label{scenarios}
\end{table}

Table \ref{scenarios} summarises the answers given by the 18 respondents. Nordhaus noted the extreme range of views, which had two polar opposites: economists ``whose principal concerns lie outside environmental economics'' and scientists, where the latter's expectations of economic damages from global warming were 30 times as large as the former's:

\begin{quote}
The major impression emerges from this survey is that experts hold vastly different views about the potential economic impact of climatic change. At one extreme are the natural scientists, \emph{all three of whom} have profound concerns about the economic impacts of greenhouse warming\ldots \emph{At the other extreme are the other subdisciplines of economics} (those whose principal concerns lie outside environmental economics); \emph{these eight respondents see much less potential for the calamitous outcome} \ldots about one-30th of the magnitude estimated by the natural scientists. \cite[p. 48. Emphasis added]{nordhaus1994expert}
\end{quote}

However, this acknowledgement was accompanied by summaries that buried this divergence in views between the climate change scientists and economists:

\begin{quote}
The second impression\ldots is that for most respondents the best guess of the impact of a 3-degree warming by 2090, in the words of respondent 17, would be ''small potatoes''. Only three respondents\footnote{We deduce, from the previous quote, that these three respondents were the three climate scientists surveyed.} expect the impact of scenario A to be more than 3 percent of GWP. In terms of economic growth, the median estimated impact for scenario A over the next century would reduce the growth of per capita incomes from, say, 1.50 percent per year to 1.485 percent per year. One respondent summarized the relaxed view: ''I am impressed with the view that \emph{it takes a very sharp pencil to see the difference between the world with and without climate change} or with and without mitigation''. \cite[p. 48. Emphasis added]{nordhaus1994expert}
\end{quote}

Despite Nordhaus's comment that ``This difference of opinion [between scientists and economists] is on the list of interesting research topics'' \cite[p. 50]{nordhaus1994expert}, this difference was not explored by any researcher in this tradition in the subsequent quarter century. A later, non-refereed literature survey on the economic impact of climate change by Nordhaus and Moffat, considered the opinions of economists only, via the economics citation database \emph{EconLit} \cite[pp. 8-10]{nordhausmoffat2017}---see \cite[p. 10]{keen2020appallingly} for further details.

\subsection{Shared data} 

``Integrated Assessment Models'' (IAMs) are used by the IPCC and policy makers to generate predictions of economic damages from climate change that then inform policy formation. F{\"u}ssel \cite[Table 1, pp. 291-92]{fussel2010modeling} lists 18 separate models, and notes the dependencies between them. Of these 18, 7 are either versions of Nordhaus's DICE, or based upon it. DICE and two other models, FUND and PAGE, were used by the USA's \emph{Interagency Working Group on Social Cost of Greenhouse Gases} \cite{interagency2016,interagency2021} to calculate the US government's estimate of the ``social cost of carbon''. These three models are calibrated on all or subsets of the numerical estimate papers outlined above. To the extent to which these models are calibrated on the flawed data points detailed here, they should not be used to assess the economic implications of climate change.

\subsection{Epistemological differences and the refereeing process}

A vital issue in this literature is how did the refereeing process lead to the publication of papers which simply assumed that activities which were not directly exposed to the weather were immune from climate change, and that today's temperature and income data could be used as a proxy for the impact of global warming on the economy? The answer may lie in a methodological principle first espoused by the influential economist Milton Friedman \cite{friedman1953methodology}, which rejected the feasibility of assessing a model on the basis of its assumptions. Given the widespread acceptance of this belief by economists,\footnote{It is, however, strongly rejected by the minority of non-Neoclassical economists, see for example \cite[Chapter 8]{keen2011debunking}.} and its incompatibility with the approach to evaluating assumptions in sciences, it is worth quoting Friedman at length:

\begin{quote}
In so far as a theory can be said to have ``assumptions'' at all, and in so far as their ``realism'' can be judged independently of the validity of predictions, the relation between the significance of a theory and the ``realism'' of its ``assumptions'' is almost the opposite of that suggested by the view under criticism. Truly important and significant hypotheses will be found to have ``assumptions'' that are wildly inaccurate descriptive representations of reality, and, in general, \emph{the more significant the theory, the more unrealistic the assumptions} (in this sense). The reason is simple. A hypothesis is important if it ``explains'' much by little, that is, if it abstracts the common and crucial elements from the mass of complex and detailed circumstances surrounding the phenomena to be explained and permits valid predictions on the basis of them alone. To be important, therefore, a hypothesis must be descriptively false in its assumptions; it takes account of, and accounts for, none of the many other attendant circumstances, since its very success shows them to be irrelevant for the phenomena to be explained. To put this point less paradoxically, the relevant question to ask about the ``assumptions'' of a theory is not whether they are descriptively ``realistic,'' for they never are, but whether they are sufficiently good approximations for the purpose in hand. And this question can be answered only by seeing whether the theory works, which means whether it yields sufficiently accurate predictions. The two supposedly independent tests thus reduce to one test. \cite[p. 153. Emphasis added]{friedman1953methodology}
\end{quote}

This approach has been strongly criticised \cite{musgrave1981unreal,ormerod1997death,archibald1963problems,mccombie2001reflections}, most cogently by the philosopher Musgrave \cite{musgrave1981unreal} , who classified assumptions into three classes---simplifying, domain and heuristic. Simplifying assumptions assert that some factor---for example, air resistance in Galileo's demonstration that two dense bodies of different weights fall at the same rate---can be neglected without significantly affecting the predictions of the theory. Domain assumptions, however, determine the range of applicability of the theory, so that if the assumption was false, then so was the model: ``\emph{The more unrealistic domain assumptions are, the less testable and hence less significant is the theory}'' \cite[p. 382]{musgrave1981unreal}.\\

Clearly, the assumptions that 87\% of industry will be unaffected by climate change because it happens in ``controlled environments'' \cite[p. 930,p. 688]{nordhaus1991slow,arent2014key}, and that ``the observed variation of economic activity with climate over space holds over time as well'' \cite[p. 32]{tol2009economic} are not simplifying assumptions, because if they are false---which they manifestly are---then the conclusions derived from them are also false. However, Neoclassical economists have ignored these critiques of Friedman's methodology, which is still widely accepted 
and taught in economics textbooks \cite[pp. 22-24]{mankiw2012principles}. Referees trained in the Neoclassical tradition, who would have reviewed Nordhaus's 1991 paper, would conceivably have treated this assumption as a ``simplifying assumption'' that was unchallengeable on methodological grounds.\\


Given the publication of this paper and several previous papers on climate-related issues \cite{nordhaus1973worlddynamics,nordhaus1974resources,nordhaus1974report,nordhaus1977economicgrowth,nordhaus1980energycrisis,nordhaus1982graze,nordhaus1991costslowing,nordhaus1991sketch}, Nordhaus himself would have frequently been called upon to referee subsequent papers on the economics of global warming. Given the tiny number of papers published in economics journals on climate change---Nicholas Stern observed in 2019 that ``the Quarterly Journal of Economics, which is currently the most-cited journal in the field of economics, has never published an article on climate change'', while the top 9 general economics journals have published 57 papers on climate change, out of a total of over 77,000 papers \cite{oswald2019economists}---this has caused a large degree of conformity in those called upon to referee economic papers on climate change. Richard Tol observed that the researchers in this field are tightly connected: ``Nordhaus and Mendelsohn are colleagues and collaborators at Yale University; at University College of London, Fankhauser, Maddison, and I all worked with David Pearce and one another, while Rehdanz was a student of Maddison and mine'' \cite[p. 31]{tol2009economic}. A refereeing process calling upon such a limited pool of researchers as referees could, as Tol also observed, lead to herding behaviour, rather than critical evaluation, which would affect the refereeing process as well as the research itself:

\begin{quote}
it is quite possible that the estimates are not independent, as there are only a relatively small number of studies, based on similar data, by authors who know each other well\ldots although the number of researchers who published marginal damage cost estimates is larger than the number of researchers who published total impact estimates, it is still a reasonably small and close-knit community who may be subject to group-think, peer pressure, and self-censoring. \cite[pp. 37, 42-43]{tol2009economic}
\end{quote}

\subsection{Post-IPCC 2014 Numerical studies}
\label{post-ipcc2014}

Studies done since the publication of the IPCC 2014 Report have produced larger estimates of the economic damages from climate change, using methodologies that differ slightly from the studies used by the IPCC. Three post-2014 papers \cite{kahn2019longterm,hsiang2017estimating,burke2015globalnonlinear} acknowledge the fallacy of using geographic temperature and income data as a proxy for the impact of global warming on income. The latest as-yet-unrefereed such study notes that:

\begin{quote}
Firstly, the literature relies primarily on the cross-sectional approach (see, for instance, Sachs and Warner 1997, Gallup et al. 1999, Nordhaus 2006, and Dell et al. 2009), and as such does not take into account the time dimension of the data (i.e., assumes that the observed relationship across countries holds over time as well). \cite[p. 2]{kahn2019longterm}
\end{quote}

However, these papers then take parameters derived from time trends between 1960 and 2014, and extrapolate them, linearly and without amendment, as predictions of what will transpire between now and 2100 at substantially higher temperatures--- 3.2$^\circ$C above today's levels, or more than 4.5$^\circ$C above pre-industrial levels, in \cite{kahn2019longterm}. This is assuming that the temperature to GDP relationships that applied between 1960 and 2014 will not be altered by increases in global temperature over the next 80 years---in other words, assuming that there are no climate tipping points in the next century, even if temperatures rise by 
more than 4$^\circ$C above pre-industrial levels. This is completely at odds with expert elicitation \cite{kriegler2009imprecise} and recent scientific evidence on tipping points \cite{lenton2019climate}.\\

These papers also maintain the assumption that only weather-exposed activities will be affected by climate change. The IPCC 1.5$^\circ$C Report \cite{masson2018} cited \cite{hsiang2017estimating} to support the assertion that the absolute decline in US GDP---that is, the difference between future GDP without climate change and future GDP with a specified increase in global average temperature---was 1.2\% per 1$^\circ$C increase over pre-industrial levels:

\begin{quote}
The economic damage in the United States from climate change is estimated to be, on average, roughly 1.2\% cost of GDP per year per 1$^\circ$C increase under RCP8.5 (Hsiang et al., 2017). \cite[p. 243]{masson2018} 
\end{quote}

These costs were the sum of primarily heat-based factors as they affected weather-affected sectors and activities only:

\begin{quote}
(A) Percent change in yields, area-weighted average for maize, wheat, soybeans, and cotton. (B) Change in all-cause mortality rates, across all age groups. (C) Change in electricity demand. (D) Change in labor supply of full-time-equivalent workers for low-risk jobs \emph{where workers are minimally exposed to outdoor temperature}. (E) Same as (D), except for high-risk jobs \emph{where workers are heavily exposed to outdoor temperatures}. (F) Change in damages from coastal storms. (G) Change in property-crime rates. (H) Change in violent-crime rates \cite[p. 3. Emphasis added]{hsiang2017estimating}. \footnote{The supplementary materials to \cite{burke2015globalnonlinear} include the statement that ``We develop empirical, micro-founded sector-specific damage functions for a number of sectors seen to be economically important. These comprise agriculture, crime, health, labor, and electricity demand.'' \cite[p. 6]{burke2015globalnonlinearsupplement}}
\end{quote}

Burke et al. generated the largest predicted damages, of a 23\% fall in GDP by 2100, by analyzing nonlinear relationships between output and temperature over the 1960-2010 period \cite{burke2015globalnonlinear}. However, they then extrapolate this forward linearly for expected temperature changes, asserting that ``If future adaptation mimics past adaptation, unmitigated warming is expected to reshape the global economy by reducing average global incomes roughly 23\% by 2100'' \cite[p. 1]{burke2015globalnonlinearsupplement}. They note that their approach ``assumes future economies respond to temperature changes similarly to today's economies---perhaps a reasonable assumption given the observed lack of adaptation during our 50-year sample'' \cite[p. 3]{burke2015globalnonlinearsupplement}. This ``reasonable assumption'' ignores the likelihood of economic damages caused by tipping points. The same applies to the unrefereed study by Kahn et al. \cite{kahn2019longterm}: their prediction of a 7.22\% fall in GWP by 2100 (consequent upon a 3.2$^\circ$C rise over 2020 temperatures) linearly extrapolates time-trends between 1960 and 2014 out to 2100.

Other research that includes in DICE the likelihood of tipping points based on scientific expert elicitation \cite{kriegler2009imprecise} leads to much higher damages and a completely different 'optimal policy' recommendation, increasing the social cost of carbon emissions more than 8-fold \cite{cailenton2016risk}.

\section{Modelling Issues} 
\label{DICE_section}

In this section we turn our attention to some key properties of the DICE model, starting with the underlying mathematical framework for the economic growth model on which it is based. We then describe how climate change is integrated in the model via the damage functions discussed in the previous sections and the associated abatement costs incurred to avoid them. The pervasive theme is that the DICE model is based on two fundamental trade-offs: the classical economics trade-off between consumption and investment and an additional trade-off between damages and abatement. The variables and parameters mentioned below are summarized in Table \ref{DICE_table}.

\subsection{Saddle-path stability of the underlying Ramsey Model}

DICE is based upon the Ramsey growth model \cite{ramsey28mathematicaltheoryofsaving}, which portrayed long-run economic growth as the product of optimal savings decisions by a highly stylised society. Since \cite{kydlandprescott1982time}, it has served as the basis of endogenous growth literature, which attempts to explain economic growth as a function of labour and population dynamics, capital accumulation or technological change. The original Ramsey model was developed in continuous time, and the model we describe below is the continuous-time version of the DICE model, using the same notation.

The key decisions in the economy are assumed to be made by a representative household,\footnote{The original Ramsey model \cite{ramsey28mathematicaltheoryofsaving} assumed an infinitely lived household that both owned and worked in the firms in the economy, maximizing its own utility by a trade-off between consumption now and consumption in the future, thus endogenizing the investment decision of firms while maximizing the consumer utility. This generated an optimal control problem of maximizing the utility function in \eqref{utility}.} which maximizes the following discounted utility 
function: 
\begin{equation}
\label{utility} 
U(c) = \int_{0}^{\infty }\frac{c(t)^{\left(1-\alpha\right)}-1}{1-\alpha} \cdot e^{-\rho t}dt.
\end{equation}
The argument of the utility function is the entire path $c = \{c(t)\}_{0\leq t\leq \infty}$ of consumption per capita 
\begin{equation}
\label{c_percapita}
    c(t) := \frac{C(t)}{L(t)},
\end{equation}
for time-varying total consumption $C(t)$ and population $L(t)$, whereas $\alpha$ is the constant relative risk aversion (CRRA) coefficient and $\rho$ is a constant discounting parameter.  

Economic output is determined by a Cobb-Douglas production function \cite{cobb1928theory} of the form
\begin{equation}
\label{CD_production}
Y(t)=A(t)L(t)^{1-\gamma}K(t)^{\gamma},
\end{equation}
where $\gamma$ is a constant parameter representing substitutability between labour $L$ and capital $K$, and $A$ is called total factor productivity (TFP)\footnote{The description of the model presented here follows the notation and definitions used in \cite{nordhaussztorc2013}. In particular, the production function in \eqref{CD_production} and \eqref{Y_available} match Equation (4) in \cite{nordhaussztorc2013}. The reader should be aware, however, that some of the calculations used in the code for the \hyperlink{https://www.gams.com/latest/docs/}{General Algebraic Modeling System} (GAMS) package used to run the model are incompatible with this description. In particular, the terminal value for the optimal savings rate matches that which would arise in a model with labour-augmenting technology where $Y(t)=(A(t)L(t))^{1-\gamma}K(t)^{\gamma}$ instead of total factor productivity where $Y(t)=A(t)L(t)^{1-\gamma}K(t)^{\gamma}$.}.

The model assumes that labour and TFP evolve exogenously.\footnote{In DICE, Nordhaus assumed that labor and TFP follow logistic time paths, with declining growth rates $g_L$ and $g_A$.} By contrast, capital evolves endogenously according to 
\begin{equation}
\frac{dK}{dt}=Y(t)-C(t)-\delta K(t),
\label{dotK}
\end{equation}
where $S = I = Y-C$ is the amount of savings that is available for investment $I$ in capital, once yearly consumption of goods and services $C$ is subtracted from total economic output. 

\begin{figure}[htbp]
	\centering
		\caption{Ramsey model. Grey boxes represent time-varying exogenous variables, blue ones represent intermediate endogenous variables, the black box corresponds to the control 	variable, and the yellow box is the objective function being optimized.}
	\includegraphics[width=10cm]{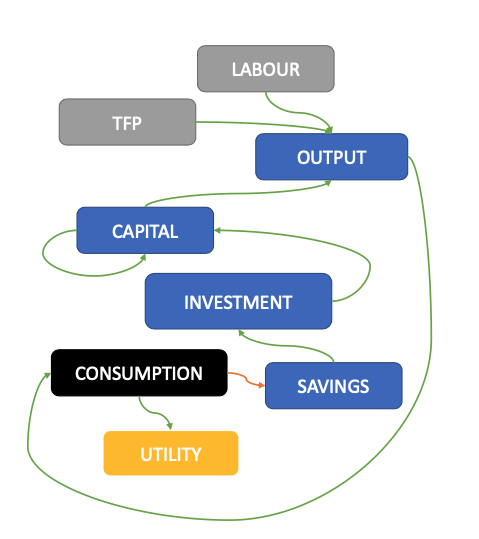}

	\label{RAMSEY}
\end{figure}

We can therefore summarize the intuition behind the model using Figure~\ref{RAMSEY} as follows: at each point in time, exogenously evolving population $L$ and TFP $A$, together with available capital stock $K$ determines production $Y$; all production is then distributed to a ``representative" household, out of which the household consumes $C$ and as a result decides on its level of savings $S$; savings are then used as investment $I$ and create new capital stock, enabling higher production in the future. 

It should be clear from this description that the representative household faces a trade-off: either consume more today and gain more utility, or save (and hence invest) to increase the capital stock and associated production capacity, and so be able to consume more and attain a higher (yet discounted) utility in the future.

This is a typical infinite horizon optimal control problem. Using standard techniques, one can express the solution for this problem in terms of a two-dimensional system of differential equations (or difference equations in the discrete-time formulation) in the two ratios consumption per capita $c$, as defined in \eqref{c_percapita}, and capital per capita, $k$:
\begin{equation}
    k(t):=\frac{K(t)}{L(t)},
\end{equation}for which one then seeks to obtain a steady state, that is to say, a stationary or equilibrium point for the variables $(c,k)$.   
As it turns out, the Jacobian at the steady state of this two-dimensional system for $(c,k)$ always has an eigenvalue with positive real part, meaning that the model is inherently unstable. What this means is that at each time $t_0$, there is only a single value $c_0$ for the consumption per capita  for which the system is able to converge to a steady state asymptotically.  

A natural question to ask, therefore, is: what happens if, because of a policy shock or some other perturbation, the economy finds itself at time $t_0$ away from the initial condition that leads to the steady state? The answer is that, in this case, the model makes the {\em additional} assumption that the variable $c$ will instantaneously jump to the value $c_0$, no matter how distant that value might be. In the Ramsey model literature, this behaviour of the ``jump variable'' $c$ is called {\em saddle path stability}.

In order to enforce this behaviour, it is sufficient to assume that at a "transversality condition" holds, namely that the solution satisfies 
\begin{equation}
\lim_{t\rightarrow \infty }U^{^{\prime }}(c)k(t)e^{-\rho t}=0  \label{e5c}
\end{equation}
In the context of constrained optimization problems with infinite horizon, this condition is the equivalent of the Kuhn-Tucker slackness condition, as it ensures a non-explosive path for the model. Any solution other than the saddle path violates this condition and is therefore discarded {\em a priori}. For a critique of this standard practice in economic models based on Ramsey, see \cite{christiaans2013economiccrises}. Christiaans notes that it ``requires extreme information on the side of economic agents who have to choose initial conditions of some jump variables with infinite accuracy'' \cite[p. 198]{christiaans2013economiccrises}. For critiques of the Ramsey model itself see \cite{kirman1989intrinsic,colander3008beyond,solow2001neoclassical,solow2007last50}.

\subsection{Incorporating emissions, damages and abatement costs}

Nordhaus's key contribution in DICE has been to couple a representation of climate dynamics with the Ramsey model in the following manner. Absent any effort to control emissions, gross output $Y_G$ is produced according to \eqref{CD_production} and generates industrial emissions $E_{ind}(t) = \sigma(t) Y_G(t)$, where $\sigma$ is the carbon intensity of the economy, which is assumed to decrease according to a logistic equation with a decreasing rate $g_\sigma$. These are added to emissions from land use, which are assumed to be exogenous emissions (i.e not subject to emission reduction efforts) $E_{exo}$, and are also assumed to decrease exponentially at a constant rate. Total emissions $E(t) = E_{ind}(t)+E_{exo}(t)$ of greenhouse gases then enter the atmospheric layer of a three-layer model and gradually diffuse through the upper and lower ocean layers. The increased concentration of greenhouse gases in the atmospheric layer, in turn, increases Earth's radiative forcing and consequently the average global temperature. As we have described in the previous sections, the increase in average global temperature compared to pre-industrial levels, $\Delta T$, is then used as the argument of a damage function $D$, which leads to output net of damages of the form 
\begin{equation}
\label{Y_net}
 Y_N(t)=(1-D(\Delta T(t))Y_G(t).
\end{equation}
In this way, damages reduce the overall quantity of production that is then allocated between consumption and future consumption through savings and investment.


The final key ingredient in the model consists of an emissions reduction rate $\mu(t)$, which is chosen by firms so that industrial emissions are reduced to 
\begin{equation}
\label{emissions}
    E_{ind}(t) = (1-\mu(t))\sigma(t) Y_G(t).
\end{equation}
Firms do so because they seek to reduce the amount paid in carbon taxes, which are assumed to be of the form 
\begin{equation}
\label{carbon_tax}
    T_C(t) = p_C(t)E_{ind}(t) = p_C(t)(1-\mu(t))\sigma(t) Y_G(t),
\end{equation}
where $p_C(t)$ is the price of carbon per ton of emissions. Reducing emissions, however, can only be achieved by incurring associated abatement costs of the form\footnote{The DICE manual \cite[pp. 12-13]{nordhaussztorc2013} states that ``The abatement cost equation is a reduced-form type model in which the costs of emissions reductions are a function of the emissions reduction rate, $\mu$. The abatement cost function assumes that abatement costs are proportional to output and to a power function of the reduction rate. The cost function is estimated to be highly convex, indicating that the marginal cost of reductions rises from zero more than linearly with the reductions rate (\ldots) The backstop technology is introduced into the model by setting the time path of the parameters in the abatement-cost equation so that the marginal cost of abatement at a control rate of 100 percent is equal to the backstop price for a given year.''}
\begin{equation}
\label{AC}
    A_C(t) = \frac{p_{BS}(t)\mu(t)^{\theta_2}}{\theta_2}\sigma(t) Y_G(t),
\end{equation}
where $p_{BS}(t)$ is the price of a backstop technology per ton of emissions, which is assumed to decreased exponentially in time, and $\theta_2$ is a constant coefficient corresponding to the convexity of costs. For a given trajectory for the carbon price $p_C(t)$, firms then choose the level of emissions reduction by minimizing the sum of carbon taxes and abatement costs,\footnote{It is straightforward to see that this corresponds to 
$\frac{\partial A_C}{\partial \mu}=-\frac{\partial T_C}{\partial \mu}$, or equivalently, the emissions reduction rate for which the marginal costs of abatement equals the marginal gain in carbon tax not paid.} namely 
\begin{equation}
\label{mu}
    \mu(t)=\min\left\{\left(\frac{p_{C}(t)}{p_{BS}(t)}\right)^{\frac{1}{\theta_2-1}},1\right\}.
\end{equation}
Once this level of emissions reduction is chosen, the corresponding abatement costs in \eqref{AC} is assumed to be further deducted from net output $Y_N$, so that the final output is given by
\begin{equation}
\label{Y_available}
    Y(t) = (1-\Lambda(t))(1-D(\Delta T(t))A(t)L(t)^{1-\gamma}K(t)^{\gamma},
\end{equation}
where $\Lambda(t) = A_C(t)/Y_N(t)$ is the ratio of abatement cost to net output. It is this output that is ultimately available for consumption and investment in \eqref{dotK} and consequently for the problem of optimizing the utility function in \eqref{utility}. 

Equivalently, for any given path for the emissions reduction rate $\mu(t)$, one can invert \eqref{mu} to obtain the carbon price trajectory $p_C(t)$ that would induce firms to choose this path, namely
\begin{equation}
\label{carbon_price}
    p_C(t) = \mu(t)^{\theta_2-1}p_{BS}(t).
\end{equation}

The fundamental assumption behind this entire block of equations is that mitigation strategies are, by construction, a cost to the economy as a whole. This assumption emerges from the literature on marginal abatement cost curves (MACC), computing the cost of reducing emissions \emph{at the sectorial level}. These MACC compare different technologies by means of unit cost of production (typically the sum the cost of machines and other types of capital, usually called \emph{CAPEX} for capital expenditures, and operating expenditures, or \emph{OPEX}, representing intermediate inputs or labour costs) and emissions. Ordering these technologies by emissions, one can then determine the marginal abatement cost from switching from a carbon-intensive technology to a less carbon-intensive one.\footnote{It is worth noting that a puzzle in this literature is the existence of negative value in these MACC, i.e. the fact that it is actually cost-efficient to use a low-carbon technology.} The issue with that hypothesis is that while it might be costly for a sector to reduce emissions, these costs are received by someone under the form of extra investment, extra employment, or extra intermediate inputs. This implies that these abatement costs are not lost for the economy as a whole, and ignoring this fact can lead to overemphasizing the cost of combating climate change. \footnote{The only way in which these costs would represent a pure loss would be the case where all factors of production (labor, capital, etc.) are fully utilised, which is certainly not a good representation of current times---though it is what is assumed in DICE.} A more subtle issue is that, when abatement costs are properly accounted as revenue for certain sectors, mitigation might lead to a rebound effect whereby the additional income leads to higher consumption and, ultimately, higher emissions. 

\subsection{On the impossibility of collapse in the DICE model}

\begin{figure}[htbp]
	\centering
		\caption{DICE model. Grey boxes represent time-varying exogenous variables, blue one for intermediate endogenous variables, black ones for control variables and the yellow box is the optimised variable.}
	\includegraphics[width=14cm]{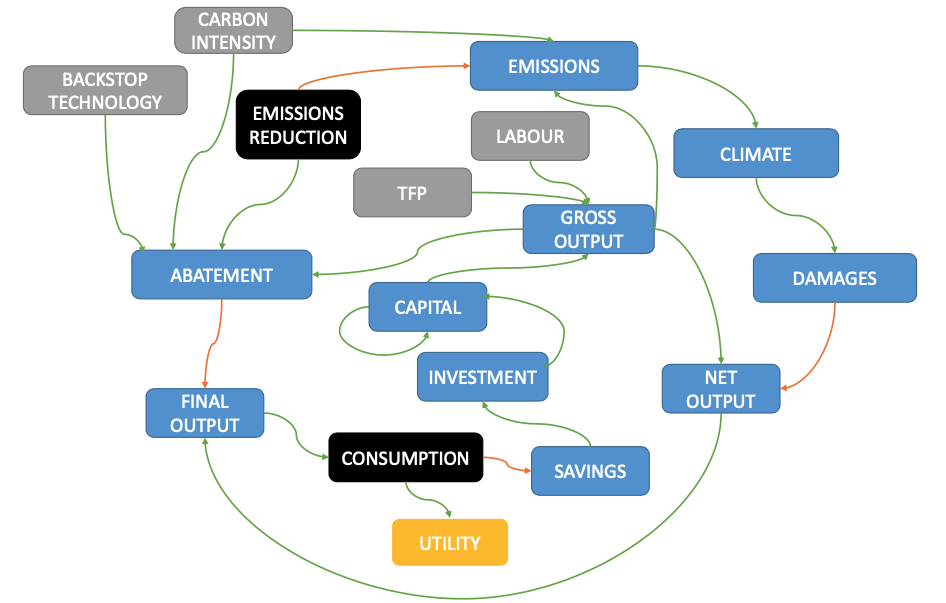}

	\label{DICE}
\end{figure}

Figure~\ref{DICE} synthesises the DICE model. When compared with the Ramsey model (see figure~\ref{RAMSEY}), one quickly sees that abatement and damages entering in the dynamics by reducing gross output to net output through damages and by further reducing net output to available output through abatement. 

Damages arise from climate change, which the model assumes to be caused by emissions, themselves depending on mitigation efforts, an exogenous carbon intensity, and gross production. Abatement is determined by gross production, mitigation effort, and costs depending on exogenous carbon intensity and backstop technology.

The fundamental difference between the Ramsey and DICE model is that there are now two control variables for optimizing the utility function: consumption as in the Ramsey model and the emissions reduction rate. Both control variables have direct and indirect effects on utility. On the one hand, consumption increases current utility but reduces future utility by reducing investment and, consequently, capital accumulation, hence leading to lower level of production in the future. On the other hand, mitigation effort reduces available production, thus affecting the optimisation between savings and consumption, but reduces emissions and, consequently, damages from climate change. Simulating DICE therefore implies determining the trajectory for both consumption/savings and emissions reduction/mitigation efforts that maximises the utility over the simulation period. 

In \cite{nordhaus2017revisiting}, Nordhaus analyses a number of different scenarios using the DICE model, among them a {\em baseline scenario}, reflecting existing efforts (or policies) to reduce climate change, and an {\em optimal scenario}, aiming to highlight the potential gains of being more ambitious in terms of mitigation policies. The baseline scenario is obtained by maximizing utility under the constraint that the carbon price remains below a given trajectory, which is specified as a simple exponential with a small growth rate chosen to 
represent existing policies projected in the future\footnote{Strictly speaking, this trajectory is the maximum between the exponential curve and {\em Hotelling price} of fossil fuel resources \cite{Hotelling1931} that is to say, the inter-temporal utility maximising price of non-renewable resources (in this case fossil fuel). This is computed by first simulating the DICE model without damages, so that the only constraint on extracting fossil fuel is related to its non-renewable resource nature: extract too much oil today and there will be not enough oil to extract it in the future. In the latest version of DICE, Nordhaus mentions that the Hotelling price is never binding, so in the end the maximum carbon price used for the baseline turns out to be simple exponential trajectory.}. The optimal scenario, by contrast, is computed by removing the maximum constraint on carbon price, allowing for a potentially higher level of utility across the entire period of simulation. Other scenarios investigated in \cite{nordhaus2017revisiting} and similar implementations of the DICE model include setting a maximum for the temperature, as well as using different values for the discounting parameter $\rho$ in \eqref{utility}.


Because our main interest in this paper is the extent to which climate damages affect the economy, we now present a different type of sensitivity analysis of the DICE model, namely with respect to the climate damage specification. Specifically, we change the value of the coefficient $a$ in the quadratic damage function \eqref{quadratic_function},\footnote{This corresponds to $a_2$ in the GAMS code used to simulate the DICE model} moving from 0.00236 (the Nordhaus specification in DICE 2017) up to a maximum value of 0.19236. This leads to increases in the damages from a maximum of 3.92\% of GDP at $4.08^\circ$C temperature increase in the Nordhaus specification to a maximum of 99.21\% of GDP at $2.27^\circ$C temperature increase. All scenarios presented hereinafter are without a cap on carbon prices, that is, corresponding to the optimal scenario of Nordhaus.

The first thing we observed is that with value $a=0.19236$, the GAMS code is not able to compute the optimal control solution and displays the error message « Infeasible solution, reduced gradient less than tolerance ». When analysing the results obtained in this case, we see that the solution does not satisfy the set of equations of the model, with capital stock jumping from a value of 223 in the first period (corresponding to 2015) to a value of 0.07 in the second period (corresponding to 2020). 

\begin{figure}[H]
\centering
\caption{Simulated gross and final economic output (left axis, trillion USD2005 per year) and damages as a proportion of gross output (right axis) in the DICE model for different specification of the damage function coefficient: Nordhaus ($a = 0.00236$), Scenario 1 ($a = 0.16236$) and Scenario 2 ($a=0.18236$).} \includegraphics [scale=0.6]{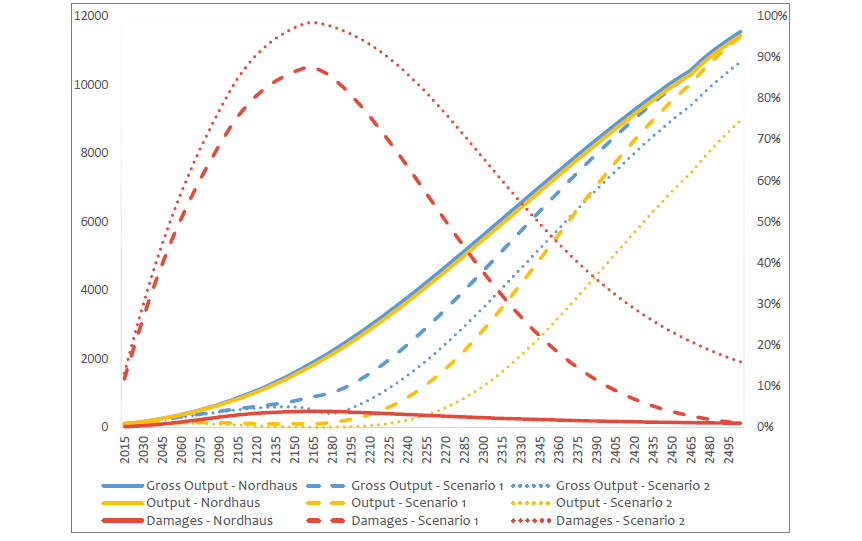}

\label{DICE_sensitivity1}
\end{figure}

\begin{figure}[H]
\centering
\caption{Simulated emissions reduction and savings rates (left axis) and abatement cost as a proportion of output net of damages (right axis) in the DICE model for different specification of the damage function coefficient: Nordhaus ($a = 0.00236$), Scenario 1 ($a = 0.16236$) and Scenario 2 ($a=0.18236$).} \includegraphics [scale=0.55]{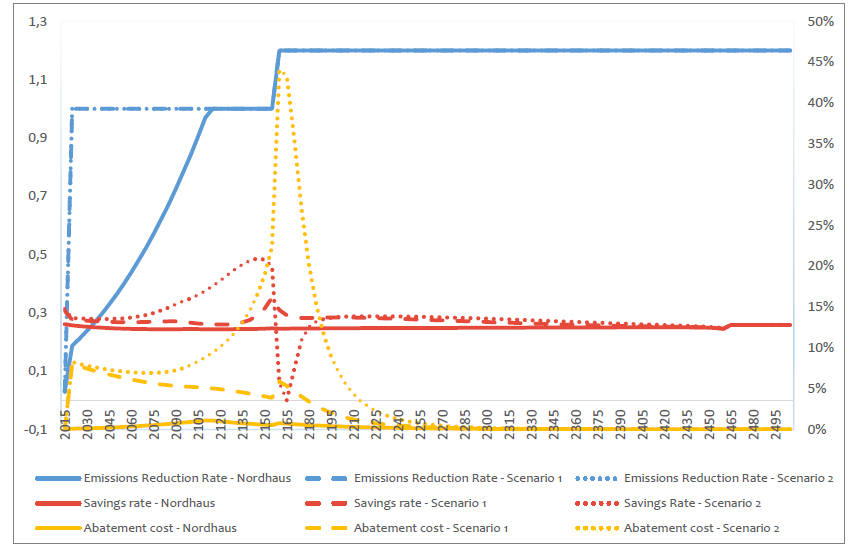}

\label{DICE_sensitivity2}
\end{figure}

\begin{figure}[H]
\centering
\caption{Simulated capital-to-output ratio (left axis, years), consumption per capita (right axis, thousand USD2005 per year), and carbon price (right axis, USD2005 per ton) in the DICE model for different specification of the damage function coefficient: Nordhaus ($a = 0.00236$), Scenario 1 ($a = 0.16236$) and Scenario 2 ($a=0.18236$).}
\includegraphics [scale=0.55]{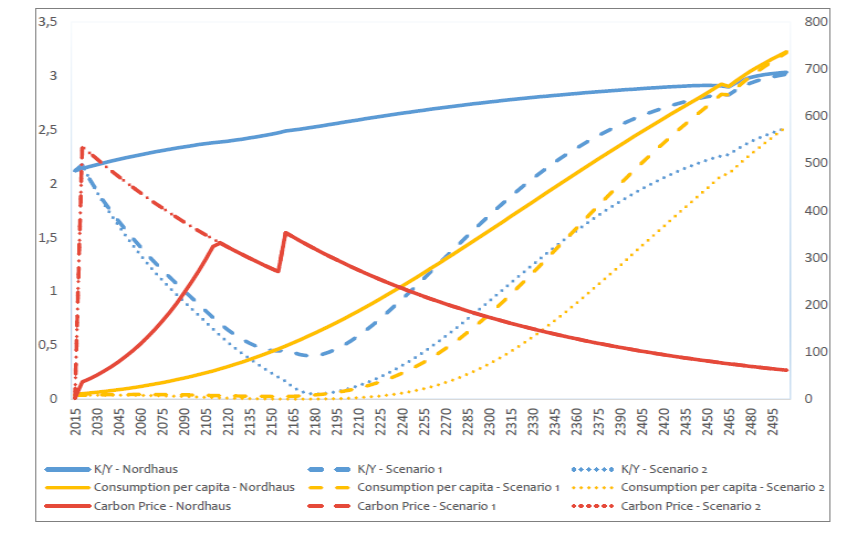}
 
\label{DICE_sensitivity3}
\end{figure}

Figures \ref{DICE_sensitivity1} to \ref{DICE_sensitivity3} present the results of the sensitivity analysis for values of $a$ equal to 0.00236 (Nordhaus -- solid lines), 0.16236 (Scenario 1 -- dashed lines) and 0.18236 
(Scenario 2 -- dotted lines). Figure \ref{DICE_sensitivity1} shows the value of gross output $Y_G$ (blue lines, left axis) defined in \eqref{CD_production}, final output $Y$ (yellow lines, left axis) as defined in \eqref{Y_available}, and damages $D(\Delta T)$ as a share of gross output (orange lines, right axis) as defined in \eqref{quadratic_function}, for the three scenarios. We first see that DICE still produces economically meaningful results, even for damages that represent more than 90\% of GDP (the maximum value for Scenario 2 being 98.46\% of GDP at $2.32^\circ$C temperature increase). 

What is more, after a period of stagnation, the economy recovers and, in the case of Scenario 1, it catches up with the Nordhaus scenario by 2500.

Figure \ref{DICE_sensitivity2} shows the two control variables: the emissions reduction rate (blue, left axis) and saving rate (red, left axis), as well as $\Lambda$, that is, abatement costs as a share of net output (purple, right axis) for the three scenarios. We observe that the two high-damages scenarios cause the emission control rate to jump immediately to its maximum value of one, while the savings rate is slightly higher. The higher emission control rate is self-explanatory. The higher savings rate is due to the fact that there are more damages, and hence investment needs to be a higher proportion of output after damages in order to lead to a desired amount of capital accumulation. The current version of DICE makes the assumption that, in 2160, the maximum level of emissions reduction rate in \eqref{mu} goes up to 1.2, corresponding to negative emissions (i.e net carbon sequestration), which in turn causes damages to decrease from this date onward as observed in Figure \ref{DICE_sensitivity1}.\footnote{We adopt this assumption in the sensitivity simulations shown here in order to stay as close to the DICE specifications as possible, but have also conducted experiments where the emissions reduction rate remains capped at one throughout the entire period. The results in this case are that damages stabilize at a constant proportion of output and there are no discontinuity in the emissions reduction rate, savings rate, and carbon price. Importantly, even though it takes longer to recover, the economy still bounces back to exponential growth in this case.} Accordingly, the years before and after this key date shows strong discontinuities in both the saving rates and the output growth rates. 
We further observe that full emission mitigation (i.e. an emissions reduction rate equal to 1) lead to to abatement costs equal to about 10\% of net output at first, whereas when the emission control rate jumps to 1.2 we observe  abatement costs amounting to more than 40\% of net output in the case of Scenario 2. 

Figure \ref{DICE_sensitivity3} shows the simulated capital-to-output ratio $K/Y$ (blue lines, left axis), consumption per capita $c$ (yellow lines, right axis) 
and the carbon price $p_C$ (red lines, right axis) for the three scenarios. In accordance with \eqref{carbon_price}, we observe that in the high-damages scenarios the optimal carbon price immediately jumps to its maximum value as the emissions reduction rate jumps to one, corresponding to the price of the backstop technology, which is assumed to decrease exponentially from a very high initial value. By contrast, in the Nordhaus scenario the optimal carbon price gradually increases up to the price of the backstop technology, at which point the emissions reduction rate is equal to 100\%. Moreover, we observe the additional discontinuity in the carbon price trajectories when the emissions reduction rate is allowed to jump to 120\% in 2160. Mirroring the behaviour of economic output observed in Figure \ref{DICE_sensitivity1}, consumption per capita remains substantially low for a long time in the high-damages scenarios---dropping below \$400 per year in the Scenario 2, possibly below subsistence levels---but eventually exhibits growth comparable to that in the Nordhaus scenario. Interestingly, the capital-to-output ratio falls to remarkably low levels in the high-damages scenarios---reaching a stunning minimum of 0.05 in Scenario 2---before bouncing back to levels comparable to those empirically observed throughout history. For comparison, the capital to output ratio for the USA was 3.36 in 2019 \cite{feenstra2015penn}, and it has never been below 1. DICE's results therefore rely upon an unprecedented, and arguably unattainable, level of capital productivity \emph{during} the period of high damages.

 Fundamentally, this sensitivity analysis highlights the strong resilience of the DICE model to increases in climate damages. When these become too high (that is, above 98\% of GDP), the model stops producing a solution.\footnote{We conjecture that changing the numerical tolerance in the GAMS package can make this threshold even higher.} Below this threshold of damages, the model still produces economically meaningful results, even if based on corner case solutions for the emission reduction rate. The cost-benefit analysis leads to full mitigation of industrial emissions (with an implied carbon price of more than 500 USD per tone) and a saving rate of around 30\%.

This extreme resilience of the economy to a collapse within the DICE model is a direct outcome of the assumptions on the functional form of the production function, full employment of the labour force and exogenously growing labour productivity. Despite higher saving rates compared to Nordhaus’s simulations, investment is lower due to high damages and high abatement costs at full emission reduction level of one and capital accumulation is much slower. However, since output depends on the population, without any unemployment dynamics, and productivity growth, which is exogenous and independent of capital accumulation, output starts growing much faster than the capital stock, and the capital-output ratio begins falling immediately after 2015. This allows for a savings rate between zero and one to keep the economy afloat without a full destruction of the capital stock until the carbon capture technology, which is assumed to kick in at 2160, starts {\em reducing} greenhouse gases concentrations and thus initializes a recovery.\footnote{As we remarked earlier, removing the assumption of net carbon capture beyond 2160 has the effect of delaying, but not preventing, a recovery in the model.} 

\subsection{Alternative Damages}

The quadratic function \eqref{quadratic_function} used by Nordhaus in the most recent version of his model is not the only type of damage function that has been proposed in the literature around the DICE model. Other economists have proposed more convex functions, notably Weitzman \cite{Weitzman2012} and Dietz and Stern \cite{DietzStern2015}. In addition to a different functional specification for the damage function itself, \cite{DietzStern2015} also consider the effects of damages to capital and to TFP, instead of damages to output only. Figure 1 in \cite{DietzStern2015} shows a comparison between a ``Standard'' model, corresponding to the unmodified DICE model, and different specifications of the damage functions, with  ``High Damage" corresponding to the Dietz and Stern functional form\footnote{The variable $S$ in the figure corresponds to the climate sensitivity parameter that affects the dynamics of the temperature increase and is assumed to be equal to 3 in the standard DICE model used for comparison. Increasing the value of this parameter to 6 as shown in the figure is meant to capture a higher level of climate risk}. As expected, introducing damages to capital and TFP lead to lower consumption per capita, as does increasing the convexity of the damage function. These results are qualitatively similar to the sensitivity results we report in Figure \ref{DICE_sensitivity3} for the consumption per capita. Observe that the simulations in \cite{DietzStern2015} stop shortly after 2200, although the same qualitative asymptotic behavior can be expected after that, namely with the consumption per capita eventually recovering from very low levels and increasing to levels comparable to those in the standard DICE model.

 

The remaining results in \cite{DietzStern2015} are also compatible with our Figures \ref{DICE_sensitivity2} and \ref{DICE_sensitivity3}. Specifically, their Tables 1 and 2 show significantly higher emissions reduction rates than in the standard DICE, in some case reaching net zero emissions about 100 years earlier, and much more rapidly increasing carbon prices. 

Two very stark conclusions arise from the results discussed in this section. First, as already remarked in \cite{DietzStern2015}, it is possible to generate much more aggressive policy recommendations within the DICE paradigm: all that is needed are realistic damage function specifications, as well as the possibility of damages affecting more than just economic output. Second, and in our view most importantly, the unique-equilibrium nature of the DICE model is incapable of generating economic collapse as a solution, {\em no matter how extreme the damages are from climate change.}

\section{Conclusion}
\label{Conclusion_section}

We conclude that there are fundamental and insurmountable weaknesses in estimates by economists of the damages from climate change, such that they should not be used to assess the risks from climate change. Any models calibrated to this data should also be rejected and not used to inform climate policy. 

The research we critique has not gone unchallenged by other economists---see for example Pindyck \cite{pindyck2013climate,pindyck2017usemisuse}, Howard and Stern \cite{howard2017few}, Weitzman \cite{weitzman2011revisiting,weitzman2011fattailed}, De Canio \cite{decanio2003economic}, Quiggin \cite{quiggin1999impact}, Ackerman \cite{ackerman2010fat,ackerman2012climate}, Nadeau \cite{nadeau2003wealth}---but the challenges have focused upon the use of high discount rates, the treatment of uncertainty about catastrophic damages, and the inappropriateness to climate change of an equilibrium modelling framework, rather than the veracity of the numerical damage estimates themselves. Unfortunately, the spurious estimates of economic damages from climate change in this literature remain the point of reference within the discipline of economics. 

They also dominate the government response to climate change to date, especially since the ``Social Cost of Carbon'' (\emph{SCC}), a concept developed by Nordhaus \cite{nordhaus2007critical,nordhaus2014estimatessocial,nordhaus2017revisiting}, has been the primary policy proposed to combat climate change. Currently, IAMs---especially DICE, PAGE and FUND---largely determine the development of government policy, to the near exclusion of the scientific literature on climate change: see in particular the publications of the USA's \emph{Interagency Working Group on Social Cost of Greenhouse Gases} \cite{interagency2016,interagency2021}, which use DICE, PAGE and FUND to calculate the \emph{SCC}, and whose academic references are dominated by economics papers, rather than the scientific literature on climate change. As De Canio observed almost two decades ago:

\begin{quote}
    The result of this modeling failure has been a bias against bold and timely action, overestimation of the cost of emissions reductions, and pervasive paralysis in the political debate \ldots the ``costs'' of climate change have been estimated as marginal displacements of consumption or GDP. The strong negative impact of the \emph{risk} of climate disruption has been systematically downplayed. \cite[p. 153]{decanio2003economic}
\end{quote}

Given the failings in the economic literature detailed in this paper, this situation cannot be allowed to continue. Where economics has failed, science should prevail. 

\bibliographystyle{RS}
\bibliography{references.bib}

\newpage

\section{Appendix}

Table \ref{table:ipccstudies} summarizes the sources for the data points in Figure \ref{fig:IPCC2014}. Table \ref{tempgdpdata} provides the data\footnote{The sources for Temperature, Population by State, and Gross State Product data are, respectively\\ \url{https://www.currentresults.com/Weather/US/average-annual-state-temperatures.php.}\\
\url{http://www2.census.gov/programs-surveys/popest/datasets/2010-2019/national/totals/}\\
\url{https://apps.bea.gov/itable/index.cfm}.\\
The data is an amalgam of average temperature by State from 1971-2000, real GSP/GDP in 2000, and population in 2010. However, similar results would apply with more consistent data. The regression result derived from it is for illustration purposes only.} used to construct Figure \ref{fig:f3gdpvtemp}. Table \ref{DICE_table} explains the variables and parameters used in DICE.

{\small
\begin{longtable}{|p{2cm}|p{1.5cm}|p{1.3cm}|p{2.0cm}|p{5cm}|}
\caption{Abbreviated from Table SM10-1 in \cite[p. SM10-4]{arent2014supplement}}
\label{table:ipccstudies}
\\ \hline
{\bf Study	}&{\bf Warming} ($^\circ$C)	& {\bf Impact (\% GDP)} &	{\bf Method}	&{\bf Coverage}\\
\hline

Nordhaus 1994 \cite{nordhaus1994managing} & 3	&-1.3&	Enumeration	&Agriculture, energy demand, sea level rise\\
\hline

Nordhaus 1994 \cite{nordhaus1994expert} &
3	&-3.6	&Expert elicitation	&Total welfare\\
\hline

Fankhauser 1995 \cite{fankhauser1995valuing} &
2.5&	-1.4	&Enumeration	&Sea level rise, biodiversity, agriculture, forestry, fisheries, electricity demand, water, resources, amenity, human health, air pollution, natural disasters\\
\hline

Tol 1995 \cite{tol1995damagecosts} &
2.5	&-1.9	&Enumeration&	Agriculture, biodiversity, sea level rise, human health, energy demand, water, resources, natural disasters, amenity\\
\hline

Nordhaus and Yang 1996 \cite{nordhaus1996regional} &
2.5	&-1.7&	Enumeration	&Agriculture, energy demand, sea level rise\\
\hline

Plambeck and Hope 1996 \cite{plambeck1996page95} &
2.5&	-2.5	&Enumeration	&Sea level rise, biodiversity, agriculture, forestry, fisheries, electricity demand, water resources, amenity, human health, air pollution, natural disasters\\
\hline

Mendelsohn 2000 \cite{mendelsohn2000country} &
2.2&	0&	Enumeration&	Agriculture, forestry, sea level rise, energy demand, water resources\\
\hline
Mendelsohn 2000 \cite{mendelsohn2000country} &
2.2&	0.1&	Statistical&	Agriculture, forestry, energy demand\\
\hline
Nordhaus and Boyer 2000 \cite{nordhaus2000warming} &
2.5&	-1.5&	Enumeration&	Agriculture, sea level rise, other market impacts, human health, amenity, biodiversity, catastrophic impacts\\
\hline
Tol 2002 \cite{tol2002estimates} &
1&	2.3&	Enumeration&	Agriculture, forestry, biodiversity, sea level rise, human health, energy demand, water resources\\
\hline
Maddison 2003 \cite{maddison2003amenity} &
2.5&	-0.1&	Statistical&	Household consumption\\
\hline
Rehdanz and Maddison 2005 \cite{rehdanz2005climate} &
1&	-0.4&	Statistical&	Self-reported happiness\\
\hline
Hope 2006 \cite{hope2006marginal} &
2.5&	-0.9&	Enumeration&	Sea level rise, biodiversity, agriculture, forestry, fisheries, energy demand, water resources, amenity, human health, air pollution, natural disasters\\
\hline
Nordhaus 2006 \cite{nordhaus2006geography} &
3&	-0.9&	Statistical&	Economic output\\
\hline
Nordhaus 2008 \cite{nordhaus2008question} &
3&	-2.6&	Enumeration&	Agriculture, sea level rise, other market impacts, human health, amenity, biodiversity, catastrophic impacts\\
\hline
Maddison and Rehdanz 2011 \cite{maddison2011impact} &
3.2&	-12.4&	Statistical&	Self-reported happiness\\
\hline
Bosello et al. 2012 \cite{bosello2012assessing} &
1.92&	-0.5&	CGE&	Energy demand; tourism; sea level rise; river floods; agriculture; forestry; human health\\
\hline
Roson and van der Mensbrugghe 2012 \cite{roson2012climate} &
2.9&	-2.1&	CGE	&Agriculture, sea level rise, water resources, tourism, energy demand, human health, labor productivity\\
\hline
Roson and van der Mensbrugghe 2012 \cite{roson2012climate} &
5.4&	-6.1&	CGE&	Agriculture, sea level rise, water resources, tourism, energy demand, human health, labor productivity\\
\hline

\end{longtable}

}

\begin{longtable}{|l|r|r|r|r|r|r|}
    \caption{USA data for state-by-state average temperature in 1971-2000, Gross State Product in 2000 and Population in 2010 used in Figure \ref{fig:f3gdpvtemp}.}
    \label{tempgdpdata}
    \\ \hline
{\bf State}&{\bf $T$ }& {\bf GSP }	& {\bf Population  } &	{\bf GSP}	&{\bf $\Delta T$}& {\bf Deviation in}\\
 &{\bf ($^\circ$C)}& {\bf (billion \$)}	& {\bf (million) } &{\bf per capita}& {\bf ($^\circ$C)} & {\bf GSP per capita}
\\ \hline

Alabama	&17.1	&119.242	&4.7797	&\$24,947&	5.6	&-\$8,259\\
\hline
Arizona	&15.7&	164.612&	6.3920&	\$25,753&	4.2&	-\$7,454\\
\hline
Arkansas&	15.8&	68.770&	2.9159&	\$23,584&	4.3&	-\$9,622\\
\hline
California&	15.2&	1,366.561&	37.2540&\$36,682&	3.7&	\$3,476\\
\hline
Colorado&	7.3&	180.606&	5.0292&	\$35,911&	-4.2&	\$2,705\\
\hline
Connecticut&	9.4&	165.899&	3.5741&	\$46,417&	-2.1&	\$13,210\\
\hline
Delaware&	12.9&	43.389&	0.8979&	\$48,321&	1.4&	\$15,115\\
\hline
Florida&	21.5&	489.488&	18.8013&	\$26,035&	10.0&	-\$7,172\\
\hline
Georgia&	17.5&	307.611&	9.6877&	\$31,753&	6.0&	-\$1,454\\
\hline
Idaho&	6.9&	37.993&	1.5676&	\$24,237&	-4.6&	-\$8,970\\
\hline
Illinois&	11.0&	487.213&	12.8306&	\$37,973&	-0.5&	\$4,766\\
\hline
Indiana&	10.9&	203.053&	6.4838&	\$31,317&	-0.6&	-\$1,890\\
\hline
Iowa&	8.8&	93.029&	3.0464&	\$30,538&	-2.7&	-\$2,669\\
\hline
Kansas&	12.4&	85.853&	2.8531&	\$30,091&	0.9&	-\$3,115\\
\hline
Kentucky&	13.1&	114.293&	4.3394&	\$26,339&	1.6&	-\$6,868\\
\hline
Louisiana&	19.1&	132.810&	4.5334&	\$29,296&	7.6&	-\$3,910\\
\hline
Maine&	5.0&	36.841&	1.3284&	\$27,734&	-6.5&	-\$5,472\\
\hline
Maryland&	12.3&	192.106&	5.7736&	\$33,274&	0.8&	\$67\\
\hline
Massachusetts&	8.8&	289.926&	6.5476&	\$44,280&	-2.7&	\$11,073\\
\hline
Michigan&	6.9&	351.573&	9.8836&	\$35,571&	-4.6&	\$2,365\\
\hline
Minnesota&	5.1&	189.965&	5.3039&	\$35,816&	-6.4&	\$2,609\\
\hline
Mississippi&	17.4&	65.646&	2.9673&	\$22,123&	5.9&	-\$11,083\\
\hline
Missouri&	12.5&	187.297&	5.9889&	\$31,274&	1.0&	-\$1,933\\
\hline
Montana&	5.9&	21.885&	0.9894&	\$22,119&	-5.6&	-\$11,087\\
\hline
Nebraska&	9.3&	56.504&	1.8263&	\$30,938&	-2.2&	-\$2,268\\
\hline
Nevada&	9.9&	76.627&	2.7006&	\$28,375&	-1.6&	-\$4,832\\
\hline
New Hampshire&	6.6&	45.226&	1.3165&	\$34,354&	-4.9&	\$1,147\\
\hline
New Jersey&	11.5&	362.007&	8.7919&	\$41,175&	0.0&	\$7,969\\
\hline
New Mexico&	11.9&	55.233&	2.0592&	\$26,823&	0.4&	-\$6,384\\
\hline
New York&	7.4&	838.660&	19.3781&	\$43,279&	-4.1&	\$10,072\\
\hline
North Carolina&	15	&275.694&	9.5354&	\$28,912&	3.5&	-\$4,294\\
\hline
North Dakota&	4.7&	17.976&	0.6726&	\$26,727&	-6.8&	-\$6,480\\
\hline
Ohio&	10.4&	391.138&	11.5365&	\$33,904&	-1.1&	\$698\\
\hline
Oklahoma&	15.3&	90.793&	3.7514&	\$24,203&	3.8&	-\$9,004\\
\hline
Oregon&	9.1&	117.258&	3.8311&	\$30,607&	-2.4&	-\$2,599\\
\hline
Pennsylvania&	9.3&	407.653&	12.7024&	\$32,093&	-2.2&	-\$1,114\\
\hline
Rhode Island&	10.1&	34.516&	1.0526&	\$32,793&	-1.4&	-\$414\\
\hline
South Carolina&	16.9&	115.247&	4.6254&	\$24,916&	5.4&	-\$8,290\\
\hline
South Dakota&	7.3&	22.691&	0.8142&	\$27,869&	-4.2&	-\$5,337\\
\hline
Tennessee&	14.2&	181.630&	6.3461&	\$28,621&	2.7&	-\$4,586\\
\hline
Texas&	18.2&	738.871&	25.1456&	\$29,384&	6.7&	-\$3,823\\
\hline
Utah&	9.2&	70.292&	2.7639&	\$25,432&	-2.3&	-\$7,774\\
\hline
Vermont&	6.1&	18.312&	0.6257&	\$29,264&	-5.4&	-\$3,942\\
\hline
Virginia&	12.8&	266.886&	8.0010&	\$33,357&	1.3&	\$150\\
\hline
Washington&	9.1&	237.832&	6.7245&	\$35,368&	-2.4&	\$2,161\\
\hline
West Virginia&	11.0&	42.607&	1.8530&	\$22,994&	-0.5&	-\$10,213\\
\hline
Wisconsin&	6.2&	180.539&	5.6870&	\$31,746&	-5.3&	-\$1,460\\
\hline
Wyoming&	5.6&	17.205&	0.5636&	\$30,526&	-5.9&	-\$2,680\\
\hline\hline
USA&	11.5&	10,252.347&	308.7455&	\$33,206&	0&	\$0\\
\hline

\end{longtable}

\begin{table}[]
\renewcommand*{\arraystretch}{1.2} 
\caption{A selection of variables and parameters in the DICE model}
\centering
\begin{tabular}{lll} 
\hline
Variable/Parameter & Related equation & Description \\
\hline
$U$ & \eqref{utility} & Discounted utility of consumption (objective function)   \\ 
$\alpha$ & \eqref{utility} & Risk aversion parameter \\
$\rho$ & \eqref{utility} & Discounting parameter \\
$C$ & \eqref{c_percapita} & Total consumption \\
$L$ & \eqref{c_percapita} & Total population and labour force (exogenous)  \\
$c$ & \eqref{c_percapita} & Consumption per capital (control)  \\
$Y$ & \eqref{CD_production} & Gross output \\
$A$ & \eqref{CD_production} & Total factor productivity (exogenous) \\
$\gamma$ & \eqref{CD_production} & Substitutability parameter  \\
$K$ & \eqref{dotK} & Total capital  \\
$\delta$ & \eqref{dotK} & Depreciation parameter \\
$I$ & \eqref{dotK}  & Investment  \\
$S$ & \eqref{dotK} & Savings \\
$Y_N$ & \eqref{Y_net} & Output net of damages \\
$E_{ind}$ & \eqref{emissions} & Industrial emissions \\
$\mu$ & \eqref{emissions}  & Emissions reduction rate (control)  \\
$\sigma$ & \eqref{emissions} & Carbon intensity (exogenous) \\
$T_C$ & \eqref{carbon_tax} & Carbon tax \\
$p_C$  & \eqref{carbon_price} & Carbon price \\
$A_C$ & \eqref{AC} & Abatement cost  \\
$p_BS$ &\eqref{AC} & Price of backstop technology (exogenous) \\
$\theta_2$ & \eqref{AC} & Convexity parameter for abatement cost \\
$Y$  & \eqref{Y_available}  & Final output  \\
$\Lambda$ & \eqref{Y_available} & Abatement cost as a fraction of net output    \\
\hline
\end{tabular}
\label{DICE_table}
\end{table}

\end{document}